\documentclass[aps,nofootinbib,groupaddress,twocolumn,floatfix,letterpaper]{revtex4}
\usepackage{graphicx}
\usepackage{amsmath}
\usepackage{dcolumn}
\usepackage{subfigure}
\usepackage{multirow}
\usepackage{mathtools}
\usepackage{epstopdf}
\usepackage{mathptmx}
\pdfoutput=1
\usepackage{color}
\definecolor{LinkColor}{rgb}{0.256,0.439,0.588}
\usepackage{hyperref}
\hypersetup{
   pdfauthor={B. D. Hauer, C. Doolin, K. S. D. Beach, J. P. Davis 
   },
   pdftitle={A general procedure for thermomechanical calibration of nano/micro-mechanical resonators},
   pdfsubject={thermomechanical calibration},
   pdfkeywords={thermomechanical,} {calibration,} {MEMS,} {NEMS,} {resonators},
   colorlinks=true,
   citecolor=LinkColor,
   linkcolor=LinkColor,
   urlcolor=LinkColor
   }

\begin{document}

\title{A general procedure for thermomechanical calibration of nano/micro-mechanical resonators}
\author{B. D. Hauer}
\email{bhauer@ualberta.ca}
\author{C. Doolin}
\author{K. S. D. Beach}
\author{J. P. Davis}
\date{\today}
\email{jdavis@ualberta.ca}
\affiliation{Department of Physics, University of Alberta, Edmonton, AB, Canada T6G 2G7}

\begin{abstract}

We provide a detailed description of a general procedure by which a nano/micro-mechanical resonator can be calibrated using its thermal motion. A brief introduction to the equations of motion for such a resonator is presented, followed by a detailed derivation of the corresponding power spectral density (PSD) function. The effective masses for a number of different resonator geometries are determined using both finite element method (FEM) modeling and analytical calculations.

\end{abstract}


\maketitle

\section{Introduction}

Nano/micro-mechanical resonators are fascinating devices, having microscopic dimensions and high quality factors ($Q$s) that allow them to be used in precise experiments. They are therefore ideal for many measurement applications, including biosensing \cite{fritz,mertens,ndieyira,gupta}, charge measurements \cite{clelandpaper}, mass sensing \cite{schmid,yang,jensen,chaste},  accelerometry \cite{krause}, temperature sensing \cite{larsen,zhang}, torque or force transduction \cite{kim,davis,gavartin}, single spin detection \cite{rugar}, and observation of qubits \cite{lahaye}. In addition, nano/micro-mechanical resonators provide an excellent platform for investigating scientific phenomena, such as vortices \cite{burgess,burgess2}, quantum zero-point motion \cite{oconnell,jchan,safavi}, molecular structure \cite{gross}, gravity \cite{weld}, superfluidity \cite{bishop}, and the Casimir force \cite{hchan}. 

As nano/micro-mechanical resonator systems become capable of performing ever more precise experiments, the sensitivity of the detection methods used to observe the devices must increase commensurately. Current state of the art techniques used for this detection include capactive measurements \cite{bay}, free-space optics \cite{suhel,rasool,fukuma}, piezoresitivity \cite{bargatin}, and optomechanics \cite{kim,anetsberger,srinivasan,eichenfield2,eichenfield3}. These methods offer extremely senstive transduction, achieving sub-attometer displacement sensitivity \cite{arcizet}, yoctogram mass resolution \cite{chaste}, and force measurements on the order of attonewtons \cite{gavartin}.

In order to quantify such sensitive measurements, it is crucial that the calibration of these devices be performed accurately. As the dimensions of nano/micro-mechanical resonators continue to decrease, this can prove to be a difficult task. However, there exists a powerful method by which the motion of these resonators can be calibrated via observation of their random thermal motion. This is known as {\it thermomechanical calibration} and is performed by invoking the equipartition theorem, which establishes a relationship between the thermal energy of a device and its mean squared amplitude of motion \cite{bowley}. Many of the detection methods mentioned above are sensitive enough to measure the thermal motion of devices on the micro or nanoscale. In addition, it has been shown that this calibration scheme agrees well with other methods, such as physically pushing on the device \cite{levy}, but it has the advantage of being noninvasive.

Thermomechanical calibration has been used to measure the motion of many systems, including cantilevers \cite{rasool,albrecht,srinivasan,higgins,rast,matei,fukuma,levy,hryciw}, doubly clamped beams \cite{hiebert,venkatesan,stievater,krause}, strings \cite{suhel,anetsberger}, torsional resonators \cite{kim,losby,clelandpaper,chabot,kleiman}, and membranes \cite{bunch,zwickl}. However, to the best of our knowledge, no single standalone document exists that describes the method of thermomechanical calibration in its entirety. We have noticed that a number of mutually inconsistent approaches to performing such a calibration can be found in the literature. This manuscript intends to elucidate the process of thermomechanical calibration by providing a full, self-consistent derivation of the procedure and ensuring that all pertinent equations are explained and defined. In this way, a general process is developed by which nano/micro-mechanical resonators can be calibrated---the only difference between varying geometries being reduced to a single parameter, the device's effective mass. In this paper, the effective mass is calculated both analytically and numerically, for a number of common device geometries. These calculations are then benchmarked against the well-known methods of Sader {\it et al.\ }\cite{sader} and Cleveland {\it et al.\ }\cite{cleveland}, which have been developed to calibrate atomic force microscopy cantilevers. In addition, we extend the methods of Sader {\it et al.\ }and Cleveland {\it et al.\ }to encompass other resonator geometries, such as doubly clamped beams and torsional resonators.

This document---which includes both a general overview of the process and specific examples---provides a clear, straightforward recipe for carrying out thermomechanical calibration for any nano/micromechanical resonator system, regardless of its complexity.

\section{Motion of a Resonator}
\label{motionsec}

We begin by describing the motion of an extended resonator structure in the linear regime. In general, the complete, three-dimensional motion will be given by a displacement function $\mathbf{R}(\mathbf{x},t)$ that can be broken down into an infinite number of independent modes. To each mode we ascribe a label, $n$, allowing for the displacement function to be expressed as \cite{salapaka,stievater}

\begin{equation}
\mathbf{R}(\mathbf{x},t) = \displaystyle\sum\limits_n a_n(t) \mathbf{r}_n(\mathbf{x}),
\label{dispeq}
\end{equation}

\noindent where $\mathbf{r}_n(\mathbf{x})$ is the mode shape of the $n^{\rm th}$ mode, and $a_n(t)$ is a function that describes the time dependence of the resonator's motion. Here we choose to normalize $\mathbf{r}_n(\mathbf{x})$ such that the maximum value of $|\mathbf{r}_n(\mathbf{x})|$ is unity. This choice of normalization ensures that $a_n(t)$ has units of distance [as the normalization leaves $\mathbf{r}_n(\mathbf{x})$ unitless] and is in 1-1 correspondence with the resonator's true physical displacement at the point of measurement. It should also be noted that by normalizing this way, the standard convention of orthonormality between the modes is lost.

Many devices can be described by simple models, for which we are able to determine $\mathbf{r}_n(\mathbf{x})$ analytically. When the device's motion is more complicated, however, we turn to FEM modeling to determine the mode shape of the structure. In this paper, these two methods are compared in a number of situations where the analytical mode shapes of the resonator can be determined.

The function $a_n(t)$ is determined by mapping each of the resonator's independent modes to a damped harmonic oscillator of the form \cite{albrecht}

\begin{equation}
\ddot{a}_n + \frac{\omega_n}{Q_n} \dot{a}_n + \omega_n^2 a_n = \frac{F(t)}{m_{\text{eff},n}}.
\label{dampharm}
\end{equation}

\noindent Here, $\omega_n$, $Q_n$, and $m_{\text{eff},n}$  are the angular frequency, quality factor, and effective mass for the $n^{\rm th}$ mode of the resonator. $F(t)$ is a time-dependent driving force. An effective spring constant for the $n^{\rm th}$ mode can also be defined and is given by $k_{\text{eff},n} = m_{\text{eff},n} \omega_n^2$ \cite{albrecht}. A general expression for $m_{\text{eff},n}$, along with its determination for a number of simple models, will be outlined in Sec.~\ref{effmasssec}.

Often a nano/micro-mechanical resonator's motion can be reduced to a one-dimensional displacement function

\begin{equation}
z(x,t) = \displaystyle\sum\limits_{n} z_n(x,t) = \displaystyle\sum\limits_{n} a_n(t) u_n(x),
\label{oned}
\end{equation}

\noindent where $u_n(x)$ describes the one-dimensional mode shape for the $n^{\rm th}$ mode of the resonator as a function of position $x$, and $a_n(t)$ is unchanged from Eq.~\eqref{dispeq}. As in the three-dimensional case, we choose to normalize $u_n(x)$ such that the maximum value of $|u_n(x)|$ is unity. This equation will be used when the displacement of a nano/micro-mechanical resonator is simple enough to be modeled by a one-dimensional function. Equations~\eqref{dispeq} and \eqref{oned} fully describe the motion of these devices in the linear regime and will be used when explicating the thermal calibration process.

\section{Power Spectral Density}
\label{psdsec}

The various methods used to detect the motion of a nano/micro-mechanical resonator share one major similarity: the motion of the resonator in question is encapsulated by some time-varying signal, typically a voltage. Generally, this measured signal is transformed into the frequency domain, providing a spectral representation of the resonator's motion that is peaked (up to small corrections in 1/$Q$) at its resonance frequency. By squaring this spectrum and dividing by the resolution bandwidth of the instrument, the measured one-sided power spectral density $S_{zz}( \omega )$ of the signal is produced \cite{bunch}.

When thermally calibrating a mechanical resonator, we must be able to relate the device's time averaged motion to its thermal energy. To do this, the equipartition theorem is used to express the device's mean-square amplitude of motion $\langle a_n^{~2}(t) \rangle$ in terms of its thermal noise PSD. In this section, we derive a theoretical expression for the PSD corresponding to a damped harmonic oscillator undergoing random thermal motion. This function can be used to properly calibrate thermal PSD data for any nano/micro-mechanical resonator.

We introduce the autocorrelation function $R_{zz}( t)$, which describes how the signal $a_n(t')$ is related to itself at a later time $t' + t$ and is defined as \cite{norton}

\begin{equation}
R_{zz}( t ) = \displaystyle  \lim_{T_0 \to \infty} \frac{1}{T_0} \int_{0}^{T_0} \! dt' \, a_n(t') a_n (t' + t).
\label{autocor}
\end{equation}

\noindent The two-sided power spectral density function $P_{zz}( \omega ) $, can then be obtained from the Fourier transform of the autocorrelation function \cite{norton}

\begin{equation}
P_{zz}( \omega ) = \int_{-\infty}^{\infty} \! dt \, R_{zz}( t ) e^{i \omega t}.
\label{PSDFour}
\end{equation}

\noindent Likewise, the autocorrelation function can be produced by performing the inverse Fourier transform

\begin{equation}
R_{zz}( t ) = \frac{1}{2 \pi} \int_{-\infty}^{\infty} \! d \omega \, P_{zz}( \omega ) e^{-i \omega t}.
\label{CorrFour}
\end{equation}

\noindent $P_{zz}( \omega)$ is defined such that it spans all frequencies, both positive and negative, and will give the total power of the signal if integrated over this entire frequency range. For real-valued signals, as is the case in physical situations, $P_{zz}(\omega)$ is an even function \cite{press}. Therefore, we introduce $S_{zz}( \omega)$, which is defined on the frequency interval from $0$ to $\infty$ such that it produces the power in the signal when integrated over all positive frequencies. From these two definitions, we find that $S_{zz}( \omega)$ and $P_{zz}( \omega)$ are related by $S_{zz} ( \omega) = 2 P_{zz}( \omega)$ \cite{press}. This relation allows the total power in the signal to be conserved when each spectral density function is integrated over the frequency interval on which it is defined. In the rest of the derivation, $S_{zz}( \omega)$ will be used, as it provides a theoretical spectral density that can be fit directly to physical data.

We define the quantity known as the mean-square amplitude of our signal as \cite{norton}

\begin{equation}
\langle {a_n}^{2}(t) \rangle = \frac{1}{T_0} \int_{0}^{T_0} \! dt \,  {[a_n(t)]}^2 .
\label{meandef}
\end{equation}

\noindent Upon inspection, we see that this quantity is equal to Eq.~\eqref{autocor} at zero time shift, provided that we have sampled the signal $a_n(t)$ over a sufficiently long time window (i.e.,~$T_0>>2 \pi / \omega_n$). Using Eq.~\eqref{CorrFour}, along with the relation between between $P_{zz}( \omega)$ and $S_{zz}( \omega)$, we can then express $\langle {a_n}^{2}(t) \rangle$ in terms of its PSD via \cite{albrecht}

\begin{equation}
\langle {a_n}^{2}(t) \rangle = \frac{1}{2 \pi} \int_{0}^{\infty} \! d \omega \, S_{zz}( \omega ).
\label{meanPSD}
\end{equation}

\noindent This result provides us with the relationship between the mean-square amplitude and PSD for $a_n(t)$. By Fourier transforming Eq.~\eqref{dampharm}, we can relate $S_{zz}( \omega)$ to $S_{FF} ( \omega )$, which is the PSD due to the forcing function $F(t)$ \cite{norton}:

\begin{equation}
S_{zz} ( \omega ) = \left| \chi( \omega ) \right|^2 S_{FF} ( \omega ).
\label{sxxsff}
\end{equation}

\noindent The factor of proportionality comes from the frequency response function \cite{krause},

\begin{equation}
\chi( \omega ) = \frac{\mathcal{F} \left\{ a_n(t) \right\}}{\mathcal{F} \left\{ F(t) \right\}} = \frac{1}{m_{\text{eff},n} \left( \omega_n^2 - \omega^2  + i \omega \omega_n / Q_n \right)}.
\label{freqres}
\end{equation}

\noindent Here $\mathcal{F} \left\{ a_n(t) \right\}$ and $\mathcal{F} \left\{ F(t) \right\}$ are the Fourier transforms of $a_n(t)$ and $F(t)$. For thermomechanical calibration, $F(t)$ is generated by random thermal noise, which is broadband in the region of interest, producing a $S_{FF}(\omega)$ that  will be flat with respect to frequency. We can therefore replace $S_{FF}(\omega)$ by a constant thermal force noise $S^{\rm th}_{FF}$. This factor can be determined by combining Eqs.~\eqref{meanPSD}--\eqref{freqres}, which results in

\begin{equation}
\langle a_n^{~2}(t) \rangle = \frac{S^{\rm th}_{FF}}{2 \pi {m^2_{\text{eff},n}}} \int_{0}^{\infty} \frac{d \omega}{ \left( \omega^2 - \omega_n^2 \right)^2 + \left( \omega \omega_n / Q_n \right)^2 }.
\label{anint}
\end{equation}

\noindent Computing this integral by contour integration, gives $\langle a_n^{~2}(t) \rangle = S^{\rm th}_{FF} Q_n / 4 \omega_n^3 {m_{\text{eff},n}}^2$ \cite{salapaka}. 

Invoking the equipartion theorem, $\langle a_n^{~2}(t) \rangle$ can be related to the thermal energy of the resonator. The equipartition theorem states that every degree of freedom that contributes a quadratic term to the total energy of a system has an average energy of $k_B T /2$ \cite{bowley}. For the time averaged potential energy of the one-dimensional harmonic oscillator considered here, this theorem is mathematically represented as \cite{higgins}

\begin{equation}
\langle U \rangle =  \frac{1}{2} m_{\text{eff},n} \omega_n^{~2} \langle a_n^2(t) \rangle = \frac{1}{2} k_B T.
\label{eqpart}
\end{equation}

\noindent It can be seen by investigation of Eq.~\eqref{dispeq} that the magnitude of $a_n(t)$ is arbitrarily determined by our choice of normalization for $\mathbf{r}_n(\mathbf{x})$. Therefore, our effective mass must be chosen very carefully to match our definition of $a_n(t)$ in order to preserve the validity of the equipartition theorem. This subtlety will be discussed in detail in Sec.~\ref{effmasssec}. 

Combining the result from Eq.~\eqref{anint} with the equipartition theorem gives \cite{albrecht}

\begin{equation}
S^{\rm th}_{FF} = \frac{4 k_B T \omega_n m_{\text{eff},n}}{Q_n},
\label{Ceq}
\end{equation}

\noindent allowing us to express the oscillator's PSD as \cite{albrecht,krause,ekinci}

\begin{equation}
S_{zz}(\omega) = \frac{4 k_B T \omega_n}{m_{\text{eff},n} Q_n\left[ \left( \omega^2 - \omega_n^2 \right)^2 + \left( \omega \omega_n / Q_n \right)^2 \right]}.
\label{PSDTang}
\end{equation}

\noindent However, since the frequency we read off our instrument is given in terms of a natural frequency, $\nu$, and not an angular frequency, it is convenient to express $S_{zz}(\omega)$ in terms of $\nu$. This is done by taking $\omega_n = 2 \pi \nu_n$, resulting in

\begin{equation}
S_{zz}(\nu) = \frac{k_B T \nu_n}{2 \pi^3 m_{\text{eff},n} Q_n \left[ \left( \nu^2 - \nu_n^2 \right)^2 + \left( \nu \nu_n / Q_n \right)^2 \right]}.
\label{PSDTfm}
\end{equation}

\noindent Alternatively, it is sometimes useful to express $S_{zz}(\nu)$ in terms of the effective spring constant $k_{\text{eff},n}$ \cite{matei,bunch,rasool,losby,fukuma}:

\begin{equation}
S_{zz}(\nu) = \frac{2 k_B T \nu_n^3}{\pi k_{\text{eff},n} Q_n\left[ \left( \nu^2 - \nu_n^2 \right)^2 + \left( \nu \nu_n / Q_n \right)^2 \right]}.
\label{PSDTfk}
\end{equation}

\noindent Equations \eqref{PSDTang}--\eqref{PSDTfk} provide the power spectrum for the thermal noise of our resonator. However, we expect a number of other sources of noise to be present in our detection system, such as detector dark current and optical shot noise \cite{hiebert}. These signals will combine with the thermal noise of the resonator to produce the power spectrum read out for the undriven device. If we assume these other sources of noise to be white, we can express our total PSD as \cite{bunch}

\begin{equation}
S_{VV}(\nu) = S_{VV}^\text{w} + \alpha S_{zz}(\nu),
\label{PSDTfit}
\end{equation}

\noindent where $S_{VV}^\text{w}$ is a constant offset to account for the white noise of the detection apparatus and $\alpha$ is a factor with units ${\rm V}^2/{\rm m}^2$ to allow for conversion of our theoretical displacement PSD (in units of ${\rm m}^2/{\rm Hz}$) to the measured voltage PSD (in units of ${\rm V}^2/{\rm Hz}$). In other words, $\alpha$ tells us how efficiently the detection system converts the motion of the resonator into the voltage signal that is measured and is closely tied to the sensitivity of the detection system. 

The experimentally measured PSD is given as $S^{\rm exp}_{VV}(\nu) = {V(\nu)}^2/\Delta \nu$ \cite{suhel} where $V(\nu)$ is the voltage spectrum of the resonator's motion obtained with a measurement bandwidth $\Delta \nu$. The experimental PSD can be fit with Eq.~\eqref{PSDTfit} near a thermal resonance using $S_{VV}^\text{w}$, $\alpha$, $\nu_n$, and $Q_n$ as fit parameters. Once $\alpha$ has been determined, it can be used to convert the PSD data $S^\text{exp}_{VV}$ from units of ${\rm V^2/Hz}$ to ${\rm m^2/Hz}$, the result of which we will call $S^\text{exp}_{xx}$. The resonator's displacement spectrum is then determined from $\displaystyle \sqrt{S^\text{exp}_{xx}}$, given in units of ${\rm m/\sqrt{Hz}}$ \cite{rasool}. From this data, we can determine the maximum displacement of the resonator, as well as the displacement sensitivity of the detection method, the latter of which is given by $\sqrt{S_{VV}^\text{w} / \alpha}$. The displacement sensitivity is often used as a figure of merit for a detection system \cite{anetsberger,srinivasan,rasool}, as it is a measure of the minimun detectable signal for that setup.

It should be noted that $\alpha$ is highly dependent on a number of factors, for instance photodiode gain or probe position, and therefore an $\alpha$ value extracted from a specific set of data should be associated only with that data set itself and not the detection system. This being said, it is possible to conceive of an experiment where all parameters are carefully controlled such that the $\alpha$ value for repeated data sets would be similar and could be combined to determine an average $\alpha$ value. Similarly, a number of data sets originating from a carefully controlled system could be combined into an averaged signal and fit with Eq.~\eqref{PSDTfit} to provide a statistically significant value of $\alpha$.
The theoretical PSD functions obtained in this section are a general result for a damped harmonic oscillator excited by random thermal fluctuations and can be used for any nano/micro-mechanical resonator. The only difference in Eq.~\eqref{PSDTfm} [Eq.~\eqref{PSDTfk}] that exists between varying resonators is the effective mass $m_{\text{eff},n}$ (spring constant $k_{\text{eff},n}$) associated with the resonator's geometry. Therefore, once we have the PSD function given by Eqs.~\eqref{PSDTfm} and \eqref{PSDTfk}, we need only focus on calculation of the device's effective mass. For this reason, the calculation of the effective mass for a number of resonator geometries will be the focus of the remaining portion of the document. For simplicity, we deal only with the quantity $m_{\text{eff},n}$, as $k_{\text{eff},n}$ is directly related to $m_{\text{eff},n}$ through the mode's experimentally determined resonance frequency.

\section{Analytical Theories of Simple Resonators}
\label{antheories}

In general, the effective mass associated with the specific mode of a resonator will depend on its mode shape. For this reason, it is useful to investigate analytical expressions for the mode shapes and resonance frequencies of some simple geometries that are commonly used for nano/micro-mechanical resonators.

\subsection{Analytical theory of beams}

A large number of nano/micro-mechanical resonators possess beam-like geometries and can be effectively modeled as macroscopic beam-like structures. The general equation of motion for a long thin beam of uniform cross section that is oscillating in one dimension is \cite{bau}

\begin{equation}
E I \frac{\partial^4 z}{\partial x^4} + \rho A \frac{\partial^2 z}{\partial t^2} = F(x),
\label{beammot}
\end{equation}

\noindent where $E$, $I$, $\rho$, and $A$ are the Young's modulus, moment of inertia, density, and cross-sectional area of the beam. $F(x)$ is an arbitrary position-dependent loading function. We know that $z(x,t)$ will be in the form of Eq.~\eqref{oned}, so we can solve Eq.~\eqref{beammot} for each of the resonator's $n$ modes separately. The oscillatory motion of the beam is encapsulated by $a_n(t)$ and, for high $Q$ resonators, is given approximately as

\begin{equation}
a_n(t) = c_n \cos(\omega_n t) + d_n  \sin(\omega_n t),
\label{beaman}
\end{equation}

\noindent where $c_n$ and $d_n$ are mode dependent constants. Inputting $z_n(x,t)$ into Eq.~\eqref{beammot} with this choice of $a_n(t)$, we obtain for the free motion [$F(x)=0$] of a beam, a differential equation for the mode shape of the $n^{\rm th}$ mode 

\begin{equation}
\frac{ d^4u_n}{dx^4} -  \frac{\lambda_n^4}{L^4} u_n = 0,
\label{beammode}
\end{equation}

\noindent where $L$ is the length of the beam and $\lambda_n = L \left( \rho A \omega_n^2 / E I \right)^{1/4}$ is a dimensionless parameter. This differential equation has the general solution \cite{bau}

\begin{equation}
\begin{split}
u_n(x) =& A_n \sin \left( \frac{\lambda_n}{L} x \right) + B_n  \cos \left( \frac{\lambda_n}{L} x \right) \\
+~& C_n \sinh \left( \frac{\lambda_n}{L} x \right) + D_n \cosh \left( \frac{\lambda_n}{L} x \right).
\end{split}
\label{beammodesol}
\end{equation}

\noindent Enforcing the beam's boundary conditions allows us to determine the values of $\lambda_n$, as well as $A_n$, $B_n$, $C_n$, and $D_n$ (up to a normalization constant), thus solving for the mode shape of each of the beam's $n$ modes. It is then possible to calculate the frequency corresponding to the $n^{\rm th}$ mode of the beam using

\begin{equation}
\nu_n = \frac{\omega_n}{2 \pi} = \frac{\lambda_n^2}{2 \pi  L^2} \sqrt{\frac{E I}{\rho A}}.
\label{beamfreq}
\end{equation}

A method of determining a cantilever's effective mass by applying a load to it and observing its deflection has been developed by Sader {\it et al.\ }\cite{sader}. It is therefore useful to determine the deformed mode shape $w(x)$ when a beam is subjected to a static loading function $F(x)$, which can be found using \cite{bau}

\begin{equation}
EI \frac{d^2u}{dx^2} = - M(x),
\label{statload}
\end{equation}

\noindent where $M(x)$ is the bending moment of the structure due to the specified loading $F(x)$. This can be solved analytically for a number of different loadings, some of which are very important in defining effective masses and will be discussed below.

Solving the above equations for the specific cases of a singly clamped beam (cantilever) and a doubly clamped beam will be very useful in providing an analytical value for the effective mass in these two simple cases. In the subsequent sections, we will treat each of these cases separately, determining both dynamic and static displacements. These will be used later to determine the effective mass of each device.

\subsubsection{Cantilevers}

The singly clamped beam, or cantilever, is subject to the boundary conditions \cite{bau}

\begin{equation}
u_n(0) = \frac{du_n}{dx}(0) = \frac{d^2u_n}{dx^2}(L) = \frac{d^3u_n}{dx^3}(L) = 0.
\label{cantbound}
\end{equation}

\begin{figure}
\includegraphics[width=\columnwidth]{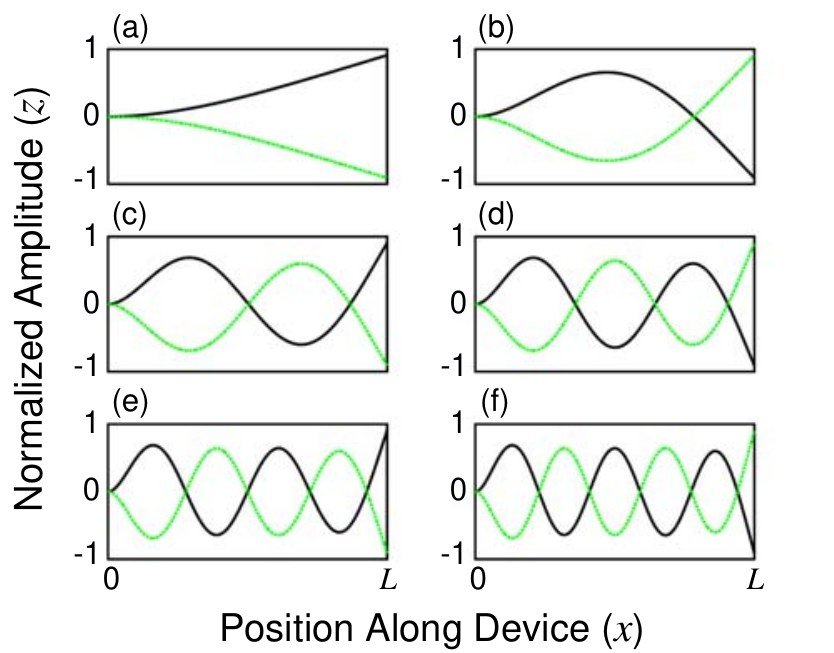}
\caption{(a)--(f) Mode shapes for the first six modes of a cantilever. The solid (black) and dashed (green) lines respresent the positive and negative maximum displacement (i.e., 180$^{\circ}$ out of phase in time with respect to each other). Each mode is normalized such that its maximum displacement is equal to one.}
\label{cantmode shapefig}
\end{figure}

\noindent These boundary conditions tell us that the cantilever is fixed and flat at $x=0$, while free at $x=L$, as can be seen in Fig.~\ref{cantmode shapefig}. Inputting these conditions into Eq.~\eqref{beammodesol}, we find that $\lambda_n$ obeys the relation \cite{clelandbook}

\begin{equation}
\cos\lambda_n \cosh\lambda_n + 1 = 0,
\label{gammacant}
\end{equation}

\noindent the solutions of which are summarized in Table \ref{lambdatab}. The mode shape of a cantilever is then determined to be \cite{salapaka}

\begin{multline}
u_n(x) = \frac{1}{K_n} \biggl( C_n \left[ \cosh\left(\frac{\lambda_n x}{L}\right) - \cos\left(\frac{\lambda_n x}{L}\right) \right] \\
- S_n \left[ \sinh\left(\frac{\lambda_n x}{L}\right) - \sin \left(\frac{\lambda_n x}{L} \right) \right] \biggr),
\label{normmodecant}
\end{multline}

\noindent where $K_n = 2 ( \sin\lambda_n \cosh\lambda_n - \cos\lambda_n \sinh\lambda_n )$, $C_n = \sinh\lambda_n + \sin\lambda_n$, and $S_n = \cosh\lambda_n + \cos\lambda_n$. Notice that our choice of normalization constant ensures that $u_n(L) = 1$. This satisfies the mode shape normalization condition found in Sec.~\ref{motionsec}, as the maximum amplitude of the cantilever is always at its free end. The first six normalized mode shapes of a cantilever can be seen in Fig.~\ref{cantmode shapefig}.

It is also useful to determine the deformed mode shape $w(x)$ of a cantilever when a static force $F_0$ is applied perpendicular to its free end [$F(x) = F_0 \delta(x-L)$], so that its effective mass can be approximated using the method of Sader {\it et al.\ } By calculating the bending moment $M(x)$ for such a loading force, we can solve for $w(x)$ using Eq.~\eqref{statload}, resulting in \cite{clelandbook}

\begin{equation}
w(x) = \frac{F_0 x^2}{2EI} \left( L - \frac{x}{3} \right).
\label{cantpointshape}
\end{equation}

\noindent The maximum displacement of the cantilever due to this loading will clearly occur at $x=L$, so that

\begin{equation}
w_{\rm max} = w(L) = \frac{F_0 L^3}{3EI}.
\label{cantpointshape}
\end{equation}

\noindent This allows us to define a corresponding spring constant

\begin{equation}
k = \frac{F_0}{w_{\rm max}} = \frac{3EI}{L^3},
\label{cantk}
\end{equation}

\noindent which will be used later to calculate the effective mass of a simple cantilever using the method of Sader {\it et al.\ }

\subsubsection{Doubly clamped beams}

The boundary conditions for a doubly clamped beam are \cite{bau}

\begin{equation}
u_n(0) = u_n(L) = \frac{du_n}{dx}(0) = \frac{du_n}{dx}(L) = 0.
\label{cantbound}
\end{equation}

\begin{figure}
\includegraphics[width=\columnwidth]{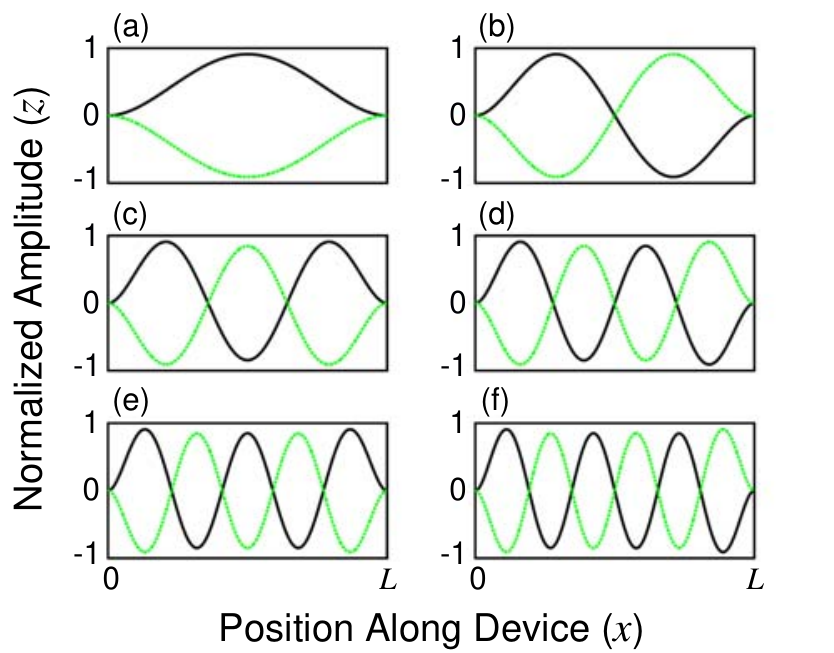}
\caption{(a)--(f) Mode shapes for the first six modes of a doubly clamped beam. The solid (black) and dashed (green) lines respresent the positive and negative maximum displacement.  Each mode is normalized such that its maximum displacement is equal to one.}
\label{beammode shapefig}
\end{figure}

\noindent Physically, this means that the beam is both flat and fixed at each of its endpoints, as can be seen in Fig.~\ref{beammode shapefig}. Inputting these conditions into Eq.~\eqref{beammodesol}, we obtain the following condition on $\lambda_n$ for doubly clamped beams \cite{clelandbook}

\begin{equation}
\cos\lambda_n \cosh\lambda_n - 1 = 0.
\label{gammabeam}
\end{equation}

\noindent The solutions to this equation are summarized in Table \ref{lambdatab}. The mode shape for a beam can also be solved and is found to have the form \cite{clelandbook}

\begin{multline}
u_n(x) =  \frac{1}{K_n} \biggl( C_n \left[ \cosh\left(\frac{\lambda_n x}{L}\right) -  \cos\left(\frac{\lambda_n x}{L}\right) \right] \\
- S_n \left[ \sinh\left(\frac{\lambda_n x}{L}\right) - \sin\left(\frac{\lambda_n x}{L}\right) \right] \biggr),
\label{modebeam}
\end{multline}

\noindent where $K_n$ is a normalization constant chosen such that the maximum value of $|u_n(x)|$ is 1, $C_n = \sinh\lambda_n - \sin\lambda_n$, and $S_n = \cosh\lambda_n - \cos\lambda_n$. It is interesting to note the similarities between Eqs.~\eqref{gammabeam} and \eqref{modebeam} for a doubly clamped beam and the corresponding Eqs.~\eqref{gammacant} and \eqref{normmodecant} for a cantilever.

Unfortunately, the position at which $|u_n(x)|$ is maximized is not obvious in the case of the doubly clamped beam, aside from the fundamental mode where the maximum is located at $x = L/2$. For this reason, the determination of $K_n$ must be carried out numerically. The first six normalized mode shapes for a simple doubly clamped beam can be seen in Fig.~\ref{beammode shapefig}.

As with the cantilever, we are also interested in the static mode shapes of doubly clamped beams due to specific loadings $F(x)$. This allows us to extend the method of Sader {\it et al.\ }to approximate the effective mass of this structure. For a point loading force $F_0$ applied perpendicularly to the center of the beam [$F(x) = F_0 \delta(x-L/2)$], Eq.~\eqref{statload} can be solved to determine the deflected mode shape $w(x)$ as \cite{clelandbook}

\begin{equation}
w(x) = \frac{F_0 x^2}{4EI} \left( \frac{L}{4} - \frac{x}{3} \right).
\label{cantpointshape}
\end{equation}

\noindent The maximum displacement occurs at $x=L/2$ and is found to be

\begin{equation}
w_{\rm max} = w(L/2) = \frac{F_0 L^3}{192EI}.
\label{cantpointshape}
\end{equation}

\noindent The corresponding spring constant $k$ due to this force is given by

\begin{equation}
k = \frac{F_0}{w_{\rm max}} = \frac{192EI}{L^3}.
\label{beamk}
\end{equation}

\begin{table}
\begin{tabular}{ ccc }
\hline \hline
Mode & \multicolumn{2}{c}{$\lambda_n$} \\  \cline{2-3}
Number ($n$) & Cantilever & Doubly Clamped Beam \\ \hline
1 & 1.8751 & 4.7300 \\
2 & 4.6941 & 7.8532\\
3 & 7.8548 & 10.9955 \\ 
$n > 3$ & $(n-1) \pi + \pi/2$ & $n \pi + \pi / 2$ \\
\hline \hline
\end{tabular}
\caption{Values of $\lambda_n$ for cantilevers and doubly clamped beams. The last row provides an approximation for $\lambda_n$ for $n>3$, which can be obtained by the fact that for large $n$, $\cosh\lambda_n$ is so large that we require $\cos\lambda_n$ to be very close to zero to satisfy Eqs.~\eqref{gammacant} and \eqref{gammabeam}.}
\label{lambdatab}
\end{table}

\subsection{Analytical theory of strings}

Another device of great interest is the nanostring \cite{suhel,verbridge,schmid,larsen}. The advantages of such structures are that they possess very high quality factors and can be modeled as macroscopic strings, which possess a very simple mode shape \cite{suhel}. The physical realization of a nanostring is to apply tension to a doubly clamped beam, which can be performed by fabricating the beam out of materials with a high intrinsic stress, such as Si$_3$N$_4$ \cite{verbridge}. Applying an axial stress in this manner will modify Eq.~\eqref{beammot} through the existence of a stress dependent term giving \cite{bau}

\begin{equation}
E I \frac{\partial^4 z}{\partial x^4} - \sigma A \frac{\partial^2 z}{\partial x^2} + \rho A \frac{\partial^2 z}{\partial t^2} = F(x),
\label{tensebeammot}
\end{equation}

\noindent where $\sigma$ is the axial stress in the beam with the axial tension in the beam due to this stress given by $N = \sigma A$. In the limit of very large tension in the beam, we can neglect the first term in Eq.~\eqref{tensebeammot} and by setting $F(x)=0$, we get the equation for the free motion of a string

\begin{equation}
\frac{\partial^2 z}{\partial x^2} - \frac{\rho}{\sigma} \frac{\partial^2 z}{\partial t^2} = 0.
\label{stringmot}
\end{equation}

\noindent This equation is solved by inputting Eq.~\eqref{oned} for $z(x,t)$ and solving for the mode shape of each of the $n$ independent modes, which are described by

\begin{equation}
\frac{\partial^2 u_n}{\partial x^2} + \frac{\rho \omega_n^2}{\sigma} u_n = 0.
\label{stringmotn}
\end{equation}

\begin{figure}
\includegraphics[width=\columnwidth]{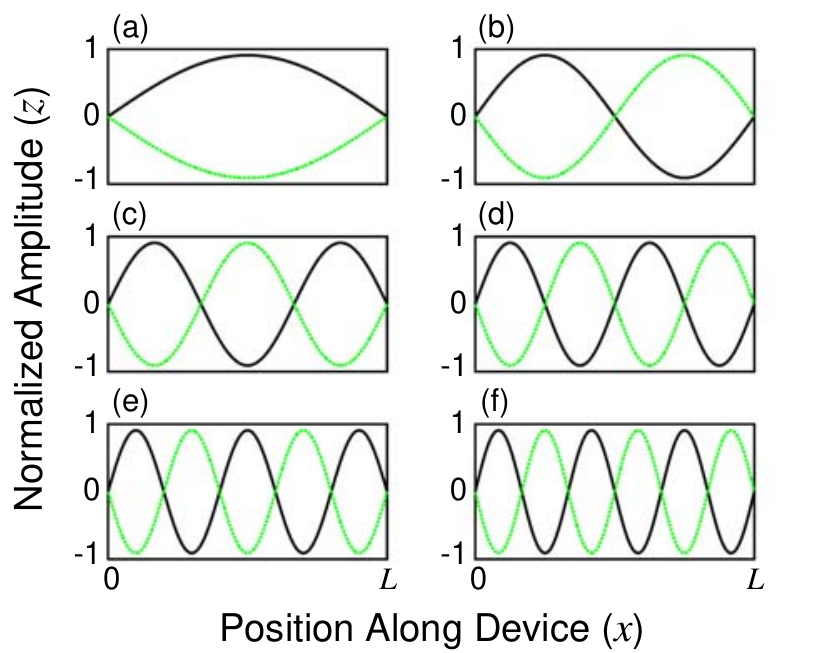}
\caption{(a)--(f) Mode shapes for the first six modes of a string. The solid (black) and dashed (green) lines respresent the positive and negative maximum displacement.  Each mode is normalized such that its maximum displacement is equal to one.}
\label{stringmodfig}
\end{figure}

\noindent Fixing the string at its endpoints $u_n(0)=u_n(L)=0$ \cite{suhel}, we get solutions \cite{schmid}

\begin{equation}
u_n(x) = \sin\left(\frac{n \pi x}{L}\right),
\label{stringmodesh}
\end{equation}

\noindent where $|u_n(x)|$ conveniently already has a maximum value of unity due to the nature of sin$(x)$, satisfying our normalization condition. The mode shapes for the first six modes of a string are shown in Fig.~\ref{stringmodfig}. The resonance frequencies of a simple string can also be determined and are given as \cite{suhel}

\begin{equation}
\nu_n = \frac{\omega_n}{2 \pi} = \frac{n}{2 L} \sqrt{\frac{\sigma}{\rho}}.
\label{stringmodef}
\end{equation}

\subsection{Analytical theory of membranes}

A membrane is one of the few two-dimensional nano/micro-mechanical systems that can be solved analytically. The physical realization of a membrane can be thought of as the two-dimensional analog of the nanostring mentioned above, in which a very thin plate is fabricated from a high stress material, such as Si$_3$N$_4$ \cite{zwickl}. Nano/micro-mechanical membranes are used in membrane-in-the-middle optomechanics experiments \cite{zwickl,flowers-jacobs,sankey} and have also been used as electrometers with charge sensitivities below the single electron level \cite{bunch}. In this section, we introduce the mode shapes and resonance frequencies of rectangular and circular membranes, which will allow us to compute their effective masses analytically in the following section.

\subsubsection{Rectangular membranes}

The equation of motion for a rectangular membrane can be obtained by generalizing the equation of motion for a string, given by Eq.~\eqref{stringmot}, to two dimensions, resulting in \cite{morse}

\begin{equation}
\frac{\partial^2 z}{\partial x^2} + \frac{\partial^2 z}{\partial y^2} - \frac{\rho}{\sigma} \frac{\partial^2 z}{\partial t^2} = 0.
\label{sqmembmot}
\end{equation}

\noindent To solve this equation, we break $z(x,y,t)$ into an infinite number of $m$ and $n$ modes in the $x$ and $y$ dimensions, respectively, giving a solution of the form

\begin{equation}
z(x,y,t) = \displaystyle\sum\limits_m \displaystyle\sum\limits_n a_{mn}(t) \psi_{mn}(x,y),
\label{twod}
\end{equation}

\noindent where $\psi_{mn}(x,y) = u_m(x) v_n(y)$ is a two-dimensional separable mode shape function that is normalized such that its maximum value is unity. Upon inspection it can be seen that this solution is the two-dimensional version of Eq.~\eqref{dispeq}, with ${\bf r}_n({\bf x}) = \psi_{mn}(x,y)$. Similar to the one-dimensional analysis, $a_{mn}(t)$ describes the oscillatory motion of the membrane and can be expressed as

\begin{equation}
a_{mn}(t) = c_{mn} \cos( \omega_{mn} t) + d_{mn} \sin(\omega_{mn} t).
\label{twod}
\end{equation}

\noindent Inputting this function for $a_{mn}(t)$ into Eq.~\eqref{sqmembmot}, we obtain an equation of motion for the mode shape of the membrane

\begin{equation}
v_n \frac{\partial^2 u_m }{\partial x^2} + u_m \frac{\partial^2 v_n }{\partial y^2} + \frac{\rho \omega_{mn}^2}{\sigma} u_m v_n = 0.
\label{sqmembmodemot}
\end{equation}

\begin{figure}
\includegraphics[width=\columnwidth]{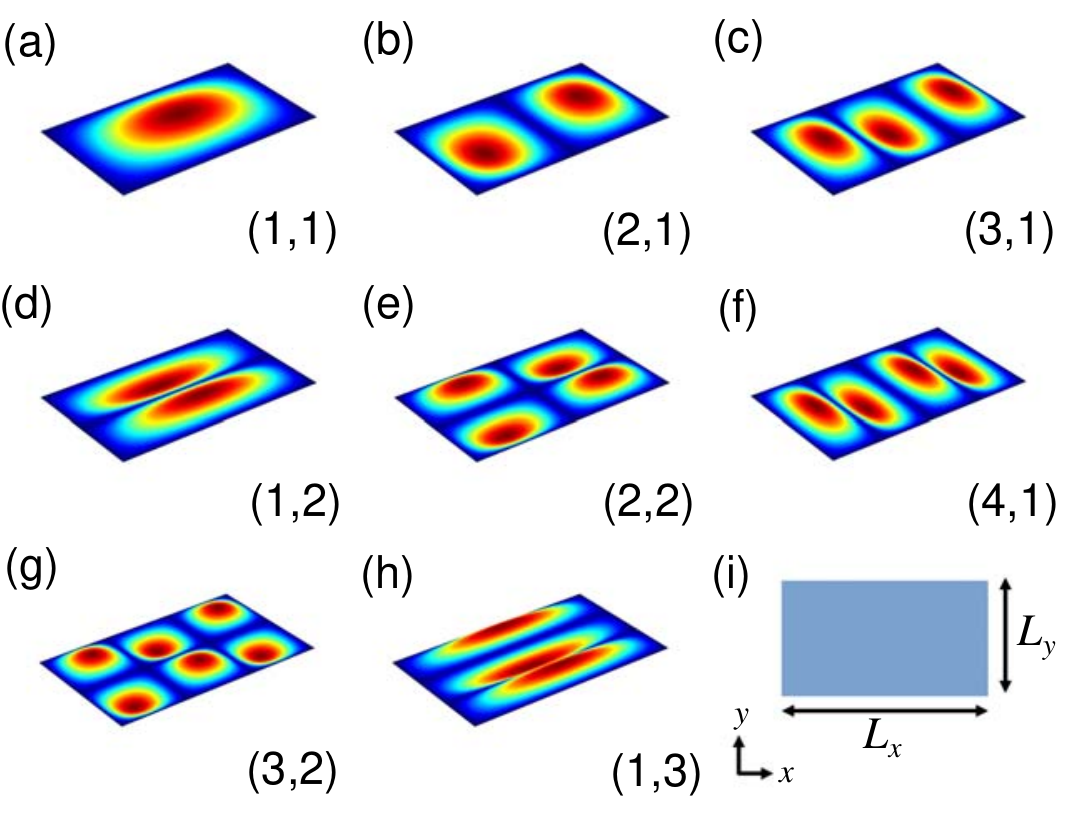}
\caption{(a)--(h) Mode shapes for the first eight modes of a rectangular membrane. Each mode is labelled with its mode index ($m$,$n$). (i) Diagram indicating the coordinates and lengths of each side of the rectangle $L_x$ and $L_y$.}
\label{rectmembmodesh}
\end{figure}

\noindent We can solve this equation using the boundary conditions for a rectangular membrane of length $L_x$ along the $x$-axis and length $L_y$ along the $y$-axis [see Fig.~\ref{rectmembmodesh}(i)] fixed at all of its edges \cite{elmore}, which are given by

\begin{equation}
\psi_{mn}(0,y) = \psi_{mn}(x,0) = \psi_{mn}(L_x,y) = \psi_{mn}(x,L_y) = 0.
\label{sqmembbound}
\end{equation}

\noindent This results in the solutions $u_m(x) = \sin (\pi m x / L_x)$ and $v_n(y) = \sin (\pi n y / L_y)$, which produce the following expression for the mode shape of the membrane \cite{morse}:

\begin{equation}
\psi_{mn}(x,y) = \sin \left(\frac{\pi m x}{L_x}\right) \sin \left(\frac{\pi n y}{L_y}\right).
\label{sqmembmodesh}
\end{equation}

\noindent Conveniently, the solution in terms of sin functions gives $\psi_{mn}(x,y)$ the normalization condition that its maximum value be unity. The first eight mode shapes for a rectangular membrane are shown in Fig.~\ref{rectmembmodesh}(a)--(h). Using this result, we can find the resonance frequency of each of the $(m,n)$ modes \cite{morse}:

\begin{equation}
\nu_{mn} = \frac{\omega_{mn}}{2 \pi} = \frac{1}{2} \sqrt{ \frac{\sigma}{\rho} \left[ \left(\frac{m}{L_x}\right)^2 + \left(\frac{n}{L_y}\right)^2\right]}.
\label{sqmembfreq}
\end{equation}

\subsubsection{Circular membranes}

Circular membranes, used in a number of experiments \cite{barton,thangawng,lindahl},  are another interesting nano/micro-mechanical resonator that have an analytically solvable mode shape.

We begin this analysis by first noting that Eq.~\eqref{sqmembmot} can be generalized to a coordinate independent form by introducing the two-dimensional Laplacian operator, $\nabla^2$, which gives \cite{morse}

\begin{equation}
\nabla^2 z - \frac{\rho}{\sigma} \frac{\partial^2 z}{\partial t^2} = 0.
\label{membmot}
\end{equation}

\noindent We note that for Cartesian coordinates, $\nabla^2 = \partial^2 / \partial x^2 + \partial^2 / \partial y^2$. The equation of motion in this form can now be solved using any system of coordinates we prefer, provided we apply an appropriate definition for $\nabla^2$.

\begin{figure}
\includegraphics[width=\columnwidth]{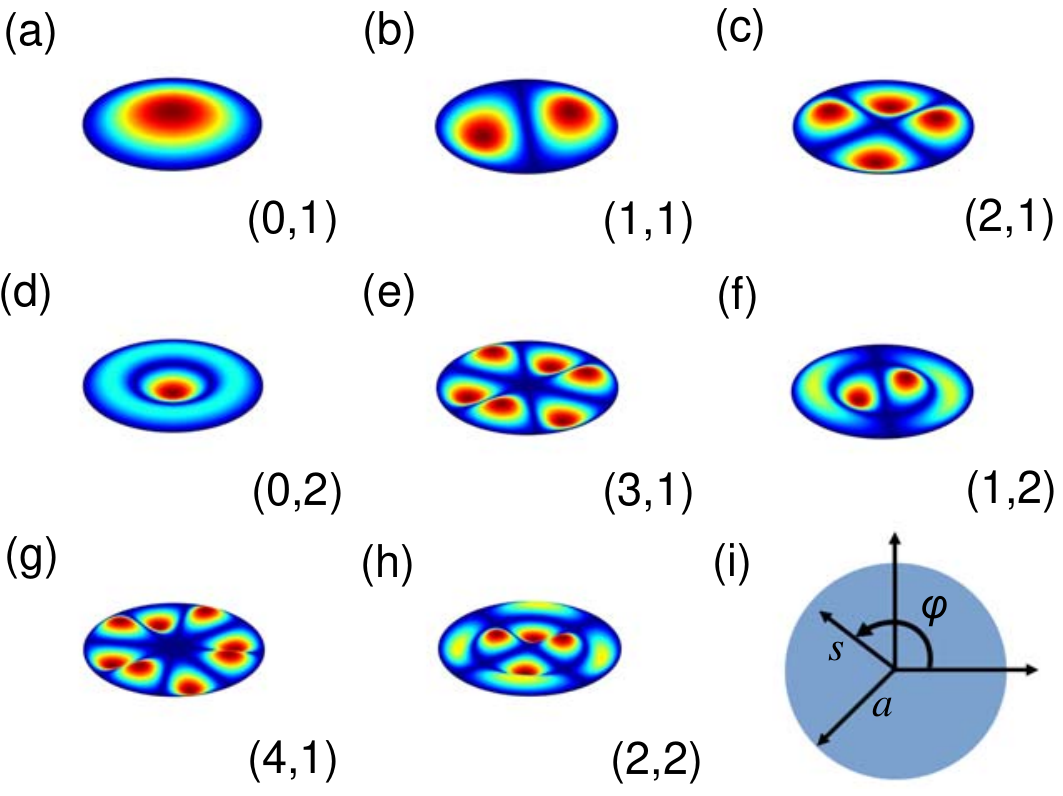}
\caption{(a)--(h) Mode shapes for the first eight modes of a circular membrane. Each mode is labelled with its mode index ($m$,$n$). (i) Diagram indicating the coordinates and radius $a$ of the circle.}
\label{circmembmodesh}
\end{figure}

When working with circular membranes, it is convenient to introduce cylindrical coordinates, as it mirrors the symmetries of the system. Transforming Eq.~\eqref{membmot} to cylindrical coordinates using the appropriate Laplacian operator gives \cite{morse}

\begin{equation}
\frac{1}{s} \frac{\partial}{\partial s} \left( s \frac{\partial z}{\partial s} \right) + \frac{1}{s^2} \frac{\partial^2 z}{\partial \phi^2}- \frac{\rho}{\sigma} \frac{\partial^2 z}{\partial t^2}= 0,
\label{circmembmot}
\end{equation}

\noindent where we have introduced the radial coordinate $s$ and the angular coordinate $\phi$ as defined in Fig.~\ref{circmembmodesh}(i). We solve this equation of motion by inputting a solution of the form

\begin{equation}
z(s,\phi,t) = \displaystyle\sum\limits_m \displaystyle\sum\limits_n a_{mn}(t) \psi_{mn}(s,\phi).
\label{twodcyl}
\end{equation}

\noindent Here $\psi_{mn}(s,\phi) = S(s) \Phi(\phi)$ is given as the product of the functions $S(s)$ and $\Phi(\phi)$, describing the membranes motion as a function of $s$ and $\phi$. We can then input Eq.~\eqref{twodcyl} into Eq.~\eqref{circmembmot}, producing the equation of motion for the mode shape of the membrane

\begin{equation}
\Phi \frac{1}{s} \frac{\partial}{\partial s} \left( s \frac{\partial S}{\partial s} \right) + S \frac{1}{s^2} \frac{\partial^2 \Phi}{\partial \phi^2} + \frac{\rho \omega_{mn}^2}{\sigma} S \Phi = 0.
\label{circmembmotmode}
\end{equation}

\noindent This equation can be solved for a fixed circular membrane of radius $a$ by ensuring that the boundary condition $\psi_{mn}(a,\phi) = 0$ is met \cite{elmore}. As well, we require the periodic condition that $\psi_{mn}(s,\phi) = \psi_{mn}(s,\phi + 2n \pi)$, which ensures that $\psi_{mn}(s,\phi)$ is single-valued.  The solutions resulting from these conditions are $\Phi(\phi) = \cos(m \phi)$ and $S(s) = J_m(\alpha_{mn} s / a)$, where $J_m$ is the Bessel function of the first kind and $\alpha_{mn}$ is determined from the solutions of $J_m(\alpha_{mn}) = 0$. Table \ref{alphatab} contains the first few values of $\alpha_{mn}$. Note that the general solution of $S(s)$ would also contain a Bessel function of the second kind, however, this function diverges as $s \rightarrow 0$, so it is discarded for physical reasons. As well, we note that $\Phi(\phi) = \sin(m\phi)$ is also a possible solution to Eq.~\eqref{circmembmotmode}, corresponding to a $\pi/2$ phase shift in $\Phi(\phi)$ meaning each mode with $m>0$ is two-fold degenerate. However, both of these degenerate modes have identical effective masses, so we concern ourselves only with the solution $\Phi(\phi) = \cos(m \phi)$ in this document. The mode shape function for a circular membrane is then given by \cite{morse}

\setlength{\tabcolsep}{5pt}

\begin{table}
\begin{tabular}{ cccccc }
\hline \hline
  \multicolumn{6}{c}{$\alpha_{mn}$} \\ 
~ & $n=1$ & $n=2$ & $n=3$ & $n=4$ & $n=5$ \\ \hline
$m=0$ & 2.4049 & 5.5201 & 8.6537 & 11.7915 & 14.9309 \\
$m=1$ & 3.8317 & 7.0156 & 10.1735 & 13.3237 & 16.4706 \\
$m=2$ & 5.1356 & 8.4172 & 11.6198 & 14.7960 & 17.9598\\
$m=3$ & 6.3802 & 9.7610 & 13.0152 & 16.2235 & 19.4094 \\
$m=4$ & 7.5883 & 11.0647 & 14.3725 & 17.6160 & 20.8269 \\
$m=5$ & 8.7715 & 12.3386 & 15.7002 & 18.9801 & 22.2178 \\
\hline \hline
\end{tabular}
\caption{Values of the $n^{th}$ zeroes for Bessel functions of the $m^{th}$ kind, $\alpha_{mn}$, for $m=0-5$ and $n=1-5$. These values are taken from \cite{tang}.}
\label{alphatab}
\end{table}

\setlength{\tabcolsep}{11pt}

\begin{equation}
\psi_{mn}(s, \phi) = K_m \cos(m \phi) J_m\left( \frac{\alpha_{mn} s}{a} \right),
\label{circmembmode}
\end{equation}

\noindent where $K_m$ is a normalization constant chosen such that the maximum value of $\psi_{mn}(s, \phi)$ is unity. The first eight modes of a circular membrane are shown in Fig.~\ref{circmembmodesh}(a)--(h). We note that due to the properties of the Bessel functions, radially symmetric modes ($m=0$) have their maximum value at $s=0$, while all other modes ($m > 0$) have a node at $s=0$ [see Fig.~\ref{circmembmodesh}(a)--(h)]. Therefore, we have that $K_0 = 1$, which is a consequence of the fact that $J_0(0) = 1$. For $m > 0$, $K_m$ is given by $\left[1 / J_m(s_{\rm max})\right]^2$, where $s_{\rm max}$ is the radial position at which $J_m(s)$ has its maximum value. To determine this value, we use the relation \cite{morse}

\begin{equation}
\frac{d J_m(s)}{ds} = \frac{1}{2} \left[ J_{m-1}(s) - J_{m+1}(s)\right] = 0,~~~ m > 0,
\label{bessderiv}
\end{equation}

\noindent which allows us to determine the extrema of $J_m(s)$ via the condition $J_{m-1}(s) = J_{m+1}(s)$. It turns out that the first solution of this equation (i.e.,~the one closest to zero), produces $s_{\rm max}$. Using this relation, the first few values of $K_m$ are determined and presented in Table \ref{kmtab}.

\begin{table}
\centering
\begin{tabular}{ cc }
\hline \hline
\multicolumn{2}{c}{$K_m$} \\ \hline
$K_0$ & 1.0000 \\
$K_1$ & 2.9536 \\
$K_2$ & 4.2251 \\
$K_3$ & 5.2995 \\
$K_4$ & 6.2609 \\
$K_5$ & 7.1456 \\
\hline \hline
\end{tabular}
\caption{Values of the normalization constant $K_m = \left[1 / J_m(s_{\rm max})\right]^2$ from Eq.~\eqref{circmembmodesh}, for $m=0-5$. }
\label{kmtab}
\end{table}

Finally, we find that the frequency for each of the modes will be given by

\begin{equation}
\nu_{mn} = \frac{\omega_{mn}}{2 \pi} = \frac{1}{2 \pi} \sqrt{\frac{\sigma}{\rho}} \frac{\alpha_{mn}}{a}.
\label{circmembfreq}
\end{equation}

We are now equipped with the necessary solutions required to analytically determine the effective masses of these simple structures. Such calculations will be investigated in the next section.

\section{Effective Mass}
\label{effmasssec}

In order to precisely calibrate a nano/micro-mechanical resonator thermomechanically, one must be careful when determining its effective mass, as numerous different methods and definitions used for calculating effective mass appear in the literature \cite{eichenfieldpaper,eichenfieldthesis,chaste,stievater,sader,cleveland,naik,dykman,groblacher,anetsberger,salapaka,albrecht,ekinci,pinard,eichenfield2,groblacher2,gigan,courty}.  What must be remembered is that we are free to use any definition of effective mass to calibrate our devices, provided that an appropriate modification is made to the measured amplitude $a_n(t)$ such that the equipartition theorem holds. In other words, in order to describe our resonators as damped harmonic oscillators, we must include a correction term in either the effective mass or the measured amplitude of the device (or both) that accounts for the extended nature of the device. However, it is imperative that one applies this correction factor self-consistently to ensure that the equipartition theorem holds. In this paper, we have chosen to define our effective mass in such a way as to leave the measured amplitude of motion uncorrected, that is to say $a_n(t)$ tracks the device's true physical motion at the position of maximum displacement. When mentioning effective mass in this paper we are referring to this definition.

The effective mass of the $n^{\rm th}$ mode of a resonator can be determined by examining its potential energy. For the $n^{\rm th}$  mode, each small element $dV$ of the resonator will have the potential energy $dU$ corresponding to that of a harmonic oscillator with a mass element $dm =\rho({\bf r})dV$ given by

\begin{equation}
dU = \frac{1}{2} \rho({\bf r}) \omega_n^2 \left|a_n(t) {\bf r}_n({\bf x}) \right|^2 dV.
\label{elepot}
\end{equation}

\noindent The total potential energy of the mode is then given by intergrating over the entire volume of the device, $V$, to get

\begin{eqnarray}
\label{pot}
U &=& \frac{1}{2} \omega_n^2 \left|a_n(t) \right|^2  \int \! dV \, \rho({\bf r}) \left| \mathbf{r}_n(\mathbf{x}) \right|^2 \\
&=& \frac{1}{2} \omega_n^2 \left|a_n(t) \right|^2 m_{\text{eff},n},
\end{eqnarray}

\noindent where we have defined \cite{eichenfieldthesis,eichenfieldpaper,naik,eichenfield2,pinard}

\begin{equation}
m_{\text{eff},n} = \int \! dV \, \rho({\bf r}) \left| \mathbf{r}_n(\mathbf{x}) \right|^2.
\label{effmass3D}
\end{equation}

\noindent With this definition, $m_{\text{eff},n}$ can be calculated once the normalized mode shape $\mathbf{r}_n(\mathbf{x})$ for the mode in question has been determined. It should be mentioned that since the PSD calculated in Sec.~\ref{psdsec} was for the time-dependent displacement function $a_n(t)$, which corresponds to the motion of the device at its point of maximum amplitude by virtue of the normalization condition for $\mathbf{r}_n(\mathbf{x})$, Eq.~\eqref{effmass3D} corresponds to the effective mass for a point-like measurement taken at this position of maximum displacement. For a point-like measurement at some other position on the resonator, $\mathbf{x}_0$, the effective mass defined in Eq.~\eqref{effmass3D} must be modified by dividing by $|\mathbf{r}_n(\mathbf{x}_0)|^2$. Therefore, the effective mass is in fact a function of the position at which we measure the device, leading to a more general definition

\begin{equation}
m_{\text{eff},n}(\mathbf{x}_0) = \frac{\int \! dV \, \rho({\bf r}) \left| \mathbf{r}_n(\mathbf{x}) \right|^2}{|\mathbf{r}_n(\mathbf{x}_0)|^2}.
\label{effmass3Dgen}
\end{equation}

Due to the fact that our normalization condition requires $|\mathbf{r}_n(\mathbf{x}_0)| < 1$, we realize that measuring the device's motion at some point $\mathbf{x}_0$ other than the point of maximum displacement results in an increase in effective mass. By inspection of Eqs.~\eqref{PSDTang}--\eqref{PSDTfk}, it can be seen that a larger effective mass results in a smaller PSD signal. It is therefore advantageous to observe a device at its position of maximum displacement, in order to increase the signal to noise ratio of the measurement. For this reason, in the remainder of the document it will be assumed that the measurement is being performed at the position of maximum displacement.

The definition given by Eq.~\eqref{effmass3D} includes devices that have a position-dependent density, such as multi-layer structures \cite{biswas} and photonic crystal devices \cite{krause,safavi}. However, most resonators have a uniform density, so we can take $\rho({\bf r}) = \rho$ in Eq.~\eqref{effmass3D}, resulting in an effective mass given by

\begin{equation}
m_{\text{eff},n} = \rho \int \! dV \, \left| \mathbf{r}_n(\mathbf{x}) \right|^2.
\label{effmass3Duni}
\end{equation}

\noindent This result can be further simplified in the case of one-dimensional motion of a device with uniform cross-sectional area $A$, in which case the effective mass will be defined as \cite{stievater}

\begin{equation}
m_{\text{eff},n} = \rho A \int_0^L \! dx \, \left| u_n(x) \right|^2.
\label{effmass1D}
\end{equation}

In this document, we collectively call Eqs.~\eqref{effmass3D}--\eqref{effmass1D} the effective mass integral (EMI). The EMI provides a general procedure for determining the effective mass of any nano/micro-mechanical resonator. As such, it is versatile and can be applied in a number of complex situations. For instance, there often exist other entities, such as supports or under-etched regions, which result from fabrication processes and move with the device itself. Such motion contributes to the energy of the mode and therefore it must be included in the effective mass calculation. A simple model describing only the motion of the device itself would therefore be an incorrect assessment of the effective mass. However, by using FEM simulation, it is possible to perform the EMI which includes the motion of all involved entities.

The EMI is also useful for structures that oscillate in more than one dimension. For most systems that can be described by simple analytical models, the motion is constrained to one dimension. However, many systems, such as membranes \cite{bunch,zwickl} and modified atomic force cantilevers \cite{reitsma}, oscillate in more than one dimension. Using FEM simulation, the EMI provides a method to determine the effective mass attributed to the complex three-dimensional motion of such devices. 

While the EMI is exact in the context of this paper, other methods for determining the effective mass for the fundamental mode of a cantilever have also been developed \cite{sader,cleveland}. One popular method introduced by Sader {\it et al.\  }is used to determine the effective mass of the first mode of a cantilever indirectly by calculating its spring constant $k$ \cite{sader}. This can be done by applying a point force to the free end of a cantilever, measuring its displacement and determining $k$ via Hooke's Law, $F = -k x$. Note that this spring constant is identical to the one defined in Eq.~\eqref{cantk}. This analysis can be performed in many ways, either analytically \cite{bau}, by simulation \cite{sader}, or experimentally \cite{levy}. Once $k$ has been determined, $m_{{\rm eff},1}$ can be calculated by the relationship $k = m_{{\rm eff},1} \omega_1^2$.

Another method of calculating the effective mass for the fundamental mode of a cantilever was developed by Cleveland {\it et al.\ }\cite{cleveland}. In this method, the effective mass is determined by adding a point mass to the free end of the cantilever, resulting in a reduced resonance frequency

\begin{equation}
\nu_1 = \frac{\omega_1}{2 \pi} = \frac{1}{2 \pi} \sqrt{\frac{k_{\text{eff},1}}{m_{{\rm eff},1}+M}},
\label{addmassBCSeq}
\end{equation}

\noindent where $M$ is the added mass. 

The two alternative methods to determine effective mass described above are developed for cantilevers. However, here we adapt these methods to use them to determine the approximate effective mass for the fundamental modes of other simple nano/micro-mechanical resonators.

It should be noted that these methods quickly break down as we move beyond the first mode or to more complex devices. It is therefore advantageous to use the EMI whenever possible when determining the effective masses of devices, as it is completely general. It should also be noted that for the definition of effective mass used in this paper, the EMI is exact and the methods of Sader {\it et al.\ }and Cleveland {\it et al.\ }are elegant approximations to this definition of effective mass. 

In the subsequent sections, we will analyze the effective mass ratio, defined as the device's effective mass $m_{\text{eff},n}$ divided by its geometric mass $m$, for the structures discussed in Sec.~\ref{antheories}. The effective mass ratio is a useful, intrinsic quantity of a system as it is independent of size or material. 

In the following calculations that involve beam-like structures, we have chosen to analyze geometries with a uniform rectangular cross section. It is important, however, to note that the results are identical for a beam with a thin, uniform cross section of any shape, provided that the structure in question can be appropriately modeled as beam-like. Justification for this claim is given by the fact that Eq.~\eqref{beammodesol} was obtained for a beam-like structure without making any assumptions other than the fact that the cross section was uniform and thin.

Using the methods described in this section, we calculate the effective mass ratio for each these structures, both analytically and through simulation. We are then able to compare the values obtained using each of the methods, allowing us to evaluate their validity.

\subsection{Effective mass of a cantilever}

We begin our calculations by computing the effective mass ratio of a cantilever analytically by inputting Eq.~\eqref{normmodecant} into the EMI, giving

\begin{equation}
\frac{m_{\text{eff},n}}{m} = \frac{1}{L} \int_0^L \! dx \, \left| u_n(x) \right|^2 = \frac{1}{4}.
\label{masscanint}
\end{equation}

\noindent This result is deceptively simple and tells us that the ratio $m_{\text{eff},n}/m$ is mode independent; that is, the effective mass of a cantilever is constant and always equal to $1/4$ of the geometric mass, regardless of mode. It should be noted that this result agrees with \cite{salapaka}, provided that the mode shape functions are normalized according to the convention presented in this paper. 

We can also use the EMI to calculate a cantilever's effective mass ratio using FEM simulation. The results for the first ten modes of a cantilever with rectangular cross section are given in Table \ref{cantmulttab}. It can be seen that these values agree well with the analytical value of $1/4$ to at least 3 decimal places. 

\begin{table}
\centering
\begin{tabular}{ cc }
\hline \hline
Mode  & $m_{\text{eff},n}/m$ \\
Label ($n$) & EMI -- FEM\\ \hline
1 & 0.2498 \\
2 & 0.2498 \\
3 & 0.2499 \\
4 & 0.2498 \\
5 & 0.2498 \\
6 & 0.2498 \\
7 & 0.2497 \\
8 & 0.2496 \\
9 & 0.2496 \\
10 & 0.2495 \\
\hline \hline
\end{tabular}
\caption{Values of $m_{\text{eff},n}/m$ calculated using FEM to solve the EMI for the first ten modes, $n$, of a cantilever with a uniform rectangular cross section.}
\label{cantmulttab}
\end{table}

We continue calculating the effective mass ratio of the fundamental mode of the cantilever using the methods of Sader {\it et al.\ }and Cleveland {\it et al.\ }We perform an analytical calculation using the method of Sader {\it et al.\ }by combining Eqs.~\eqref{beamfreq} and \eqref{cantk}, which can be rearranged to the form $\omega_1 = \sqrt{k/m_{{\rm eff},1}}$, thus identifying an effective mass. The result is

\begin{equation}
\omega_1 = \sqrt{\frac{E I{\lambda_1}^4}{\rho AL^4}} = \sqrt{ \frac{k {\lambda_1}^4}{3m}} = \sqrt{\frac{k}{m_{{\rm eff},1}}},
\label{masscansad}
\end{equation}

\noindent where we have taken $m_{{\rm eff},1} = 3m / {\lambda_1}^4$ as our effective mass. Using the value of $\lambda_1$ found in Table \ref{lambdatab} for a cantilever, we find $m_{{\rm eff},1} /m = 0.2427$ for this method. It should be reiterated that this method of determining the effective mass of a cantilever is limited to the fundamental mode. This can easily be seen by the fact that if we tried to extend the definition to the $n^{\rm th}$ mode, the result would be $m_{\text{eff},n}/m = 3 / \lambda_n^4$, which incorrectly produces a rapidly decreasing effective mass ratio as $n$ increases.  This is in direct contradiction to Eq.~\eqref{masscanint}, which indicates that the effective mass ratio for a cantilever is constant. The reason why this method cannot be extended beyond the fundamental mode is that the spring constant $k$ we calculate here is only meaningful for the first mode, i.e., the spring constant is mode dependent with our definition of effective mass.

The effective mass ratio can also be calculated using FEM simulation for this method. A force is applied to the free end of a cantilever and, using stationary analysis, $k$ is calculated. An eigenfrequency solver is then used to obtain the frequency of the fundamental mode of the same cantilever. The effective mass is then determined from $m_{{\rm eff},1} = k / (2 \pi \nu_1)^2$. Using this method we find that $m_{{\rm eff},1}/m=0.2425$.

\begin{figure}
\includegraphics[width=\columnwidth]{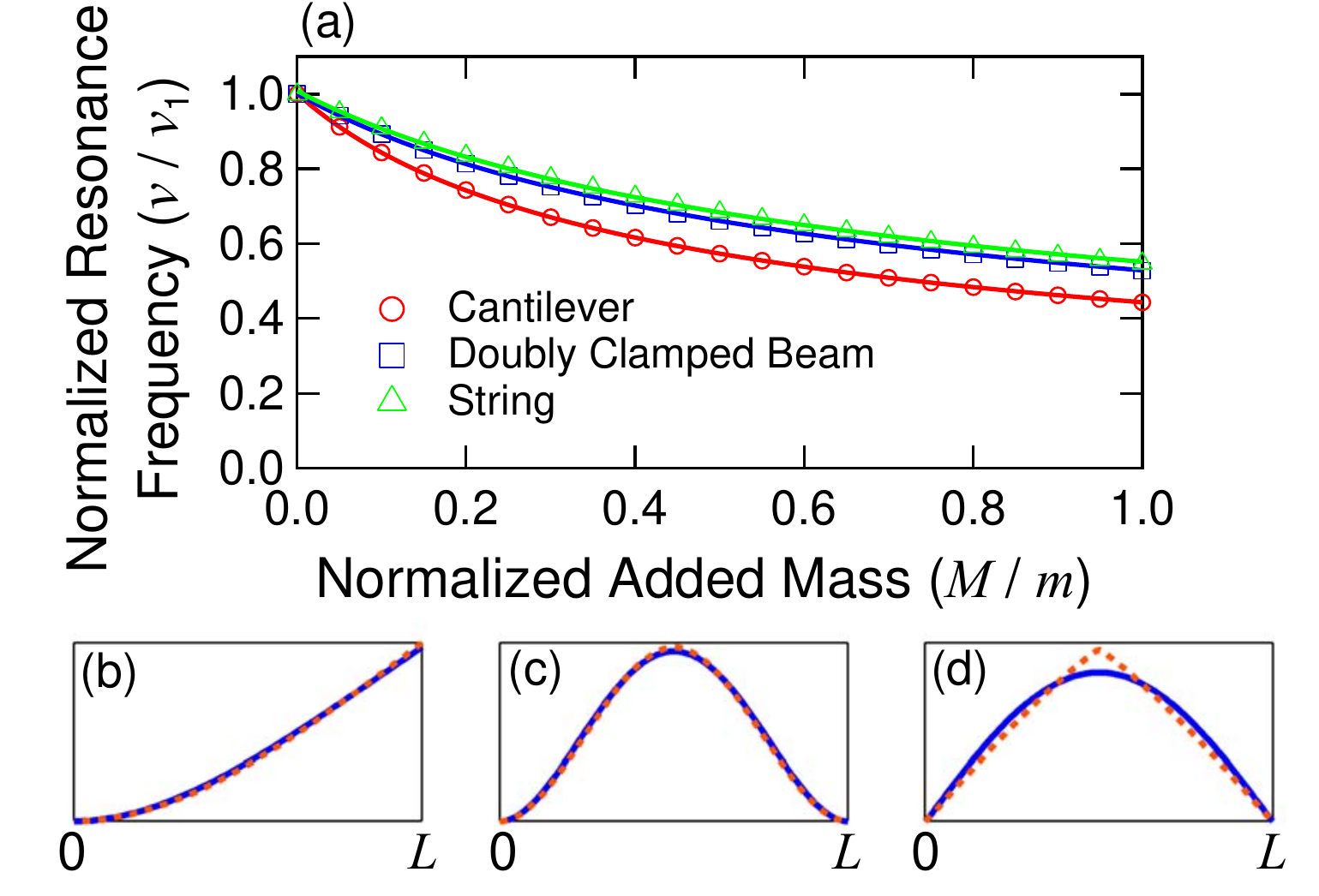}
\caption{(a) Plot of normalized frequency versus normalized added mass for the fundamental modes of a cantilevers (red circles), doubly clamped beams (blue squares) and strings (green triangles). The frequency values are normalized by dividing by the fundamental frequency of each device with $M=0$. The added mass is normalized by dividing by the total geometric mass $m$ of the device. Solid lines are fits to Eq.~\eqref{addmassBCSeq}. Effective mass ratios extracted from the fits are $m_{{\rm eff},1}/m = 0.2438,~ 0.3864,~0.4265$ for the cantilever, doubly clamped beam and string, which can be found in Tables \ref{cantfundtab}, \ref{beamfundtab}, and \ref{stringfundtab}, respectively. (b) FEM mode shape for the fundamental mode of a cantilever for both the unloaded ($M=0$: solid blue) and maximally loaded ($M=m$: dashed orange) cases. Similar plots for a (c) beam and (d) string, are also presented.}
\label{addmassBCSfig}
\end{figure}

Finally, we can calculate the effective mass ratio for a cantilever using the method of Cleveland {\it et al.\ } This is performed using FEM simulation, in which a point mass is added to the end of the cantilever and the corresponding change in resonance frequency is observed. This analysis is performed with additional masses of varying magnitudes, producing the data points that are plotted in Fig. \ref{addmassBCSfig}(a). This data is then fit with Eq. \eqref{addmassBCSeq}, from which we can extract the effective mass ratio. For the cantilever data, we find that $m_{{\rm eff},1} /m = 0.2438$ using this method.

Again, this method cannot be extended past the fundamental mode. The reason in this case is that the added mass at the free end, while not adversely affecting the first mode, significantly modifies the mode shape of higher order modes, preventing an accurate determination of the effective mass in these cases.

It can be seen that the methods of Sader {\it et al.\ }and Cleveland {\it et al.\ }provide a very good approximation to the effective mass for the fundamental mode of a simple cantilever. This is made evident by comparing $m_{{\rm eff},1} / m$ for the varying methods presented here, which are summarized in Table \ref{cantfundtab}.

It should be mentioned that the effective mass calculations performed in this section were for a simple cantilever with uniform rectangular cross section, allowing us to solve for its effective mass analytically. In general, however, the geometry of a cantilever will be more complex, requiring the use of FEM simulation to calculate its effective mass.

\begin{table}
\centering
\begin{tabular}{ lc }
\hline \hline
Method & $m_{{\rm eff},1}/m$ \\ \hline
EMI -- Analytical & 0.2500 \\
EMI -- FEM & 0.2498 \\
Sader {\it et al.\ }-- Analytical & 0.2427 \\
Sader {\it et al.\ }-- FEM & 0.2425 \\
Cleveland {\it et al.\ }-- FEM & 0.2438 \\
\hline \hline
\end{tabular}
\caption{Values of $m_{\text{eff},n}/m$ for the fundamental mode of a simple cantilever calculated using the different methods detailed in Sec.~\ref{effmasssec}.}
\label{cantfundtab}
\end{table}

\subsection{Effective mass of a doubly clamped beam}

Just as with the cantilever, we will begin the effective mass ratio calculations for a doubly clamped beam by computing the analytical result obtained using Eqs.~\eqref{modebeam} and \eqref{effmass1D}:

\begin{equation}
\frac{m_{\text{eff},n}}{m} = \frac{1}{L} \int_0^L \! dx \, | u_n(x) |^2.
\label{massstrint}
\end{equation}

\noindent Unfortunately, the ratio $m_{\text{eff},n}/m$ is not mode independent for a doubly clamped beam as it is for a cantilever. However, it appears that for $n\geq4$ it saturates to a value of $m_{\text{eff},n} / m \approx 0.4372$. We can also compute the effective mass ratio by solving the EMI using FEM simulation as we did with the cantilever. Both the analytical and FEM results are given for the first ten modes of a simple doubly clamped beam in Table \ref{beammulttab}.

\begin{table}
\centering
\begin{tabular}{ ccc }
\hline \hline
Mode & \multicolumn{2}{c}{$m_{\text{eff},n}/m$} \\ \cline{2-3}
Label ($n$) & EMI -- Analytical & EMI -- FEM \\ \hline
1 & 0.3965 & 0.3959\\
2 & 0.4390 & 0.4381\\
3 & 0.4371 & 0.4358\\
4 & 0.4372 & 0.4353\\
5 & 0.4372 & 0.4347\\
6 & 0.4372 & 0.4339\\
7 & 0.4372 & 0.4330\\
8 & 0.4372 & 0.4320\\
9 & 0.4372 & 0.4309\\
10 & 0.4372 & 0.4296\\
\hline \hline
\end{tabular}
\caption{Values of $m_{\text{eff},n}/m$ calculated by solving the EMI both analytically and using FEM for the first ten modes, $n$, of a doubly clamped beam with uniform rectangular cross section.}
\label{beammulttab}
\end{table}

\setlength{\tabcolsep}{11pt}

The method of Sader {\it et al.\ }was originally developed for a cantilever. However, we can imagine an analogous method for doubly clamped beams in which we apply a point force in the middle of the beam instead of at the end, as that is where we expect to have the maximum displacement for the fundamental mode. We can find an analytical value for $m_{{\rm eff},1} / m$ by combining Eqs.~\eqref{beamfreq} and \eqref{beamk}, which results in

\begin{equation}
\omega_1 = \sqrt{ \frac{ k {\lambda_1}^4}{192m}} = \sqrt{\frac{k}{m_{{\rm eff},1}}}.
\label{masscansad}
\end{equation}

\noindent Here we identify $m_{{\rm eff},1} = 192 m / {\lambda_1}^4$ as the effective mass of the doubly clamped beam in this equation. Using the value of $\lambda_1$ for a doubly clamped beam from Table \ref{lambdatab}, we find that $m_{{\rm eff},1} / m = 0.3836 $. We can also perform a similar analysis using FEM, which produces $m_{{\rm eff},1} / m = 0.3830$. Both of these results can be found in Table \ref{beamfundtab}.

We can also extend the method of Cleveland {\it et al.\ }for determining the effective mass of a cantilever to an analogous process for a doubly clamped beam. Again, instead of applying the mass loading to the free end as with the cantilever, the mass is added to the center of the beam. Such a loading perturbs the fundamental mode of the beam by a small amount [see Fig.~\ref{addmassBCSfig}(c)]. By varying this mass and monitoring the resulting change in resonance frequency, we produce the curve seen in Fig.~\ref{addmassBCSfig}(a). Extracting an effective mass ratio from the fit to this data gives a value of $m_{{\rm eff},1} / m = 0.3864$.

By examining Table \ref{beamfundtab}, we can see that the methods of Sader {\it et al.\ }and Cleveland {\it et al.\  }adapted for a doubly clamped beam provide very good approximations to the effective mass of the fundamental mode. This agreement suggests that making such an analogy is reasonable for this simple geometry.

\begin{table}
\begin{tabular}{ lc }
\hline \hline
Method & $m_{{\rm eff},1}/m$ \\ \hline
EMI -- Analytical & 0.3965 \\
EMI -- FEM & 0.3959 \\
Sader {\it et al.\ }-- Analytical & 0.3836 \\
Sader {\it et al.\ }-- FEM & 0.3830 \\
Cleveland {\it et al.\ }-- FEM & 0.3864 \\
\hline \hline
\end{tabular}
\caption{Values of $m_{\text{eff},n}/m$ for the fundamental mode of a simple doubly clamped beam calculated using the different methods detailed in Sec.~\ref{effmasssec}.}
\label{beamfundtab}
\end{table}

\subsection{Effective mass of a string}

As the mode shape for a string is very simple, calculation of its effective mass ratio by inputting Eq.~\eqref{stringmodesh} into the EMI is straightforward and gives \cite{suhel}

\begin{equation}
\frac{m_{\text{eff},n}}{m} = \frac{1}{L} \int_0^L \! dx \, \left| u_n(x) \right|^2 = \frac{1}{2}.
\label{massstrint}
\end{equation}

\noindent Similar to the cantilever case, the effective mass is mode independent. The analytical result above can also be seen to agree well with FEM solutions of the EMI, which are summarized in Table \ref{stringmulttab}.

\begin{table}
\centering
\begin{tabular}{ cc }
\hline \hline
Mode & $m_{\text{eff},n}/m$ \\ 
Label ($n$) & EMI -- FEM\\ \hline
1 & 0.4969 \\
2 & 0.4966 \\
3 & 0.4961 \\
4 & 0.4955 \\
5 & 0.4946 \\
6 & 0.4933 \\
7 & 0.4908 \\
8 & 0.4911 \\
9 & 0.4898 \\
10 & 0.4883 \\
\hline \hline
\end{tabular}
\caption{Values of $m_{\text{eff},n}/m$ calculated by solving the EMI using FEM for the first ten modes, $n$, of a simple string.}
\label{stringmulttab}
\end{table}

As with the doubly clamped beam, we can attempt to adapt the methods of Sader {\it et al.\ }and Cleveland {\it et al.\ }to determine approximations to the effective mass ratio of a string.

Due to its geometry, it is very difficult to conceive of an analytical method by which one could calculate the spring constant of a string. However, it is possible to imagine applying the method of Sader {\it et al.\ }to a beam with an axial tension, which mimics a string at large tension values. We attempt to determine the effective mass of a string in this way, with the aid of FEM simulation. It should be noted that one must be very careful when applying such a force using FEM simulation, as nonlinear effects must be considered in this case. Using this method, a value of $m_{{\rm eff},1} / m = 0.4334$ is found. 

Using the method of Cleveland {\it et al.,\ }in which an added mass is applied at the center of the string, FEM simulation produces the data seen in Fig.~\ref{addmassBCSfig}. Fitting this curve with Eq.~\eqref{addmassBCSeq} gives a result of $m_{{\rm eff},1} / m = 0.4265$. 

The values found using the methods of Sader {\it et al.\ }and Cleveland {\it et al.,\ }while similar to each other, deviate significantly from the analytical value of $1/2$ we found in Eq.~\eqref{massstrint}. This demonstrates a much greater discrepancy between exact and approximate values than what was observed when applying similar methods to doubly clamped beams and cantilevers. This is due to the fact that the mode shape of the string is much more senstive to added mass, which is likely a result of the high tension in the device. This hypothesis is confirmed by observation of FEM mode shapes, as can be seen in Fig.~\ref{addmassBCSfig}(d), and by the fact that if we fit a segment of the curve which excludes larger masses, we get a value that is closer to the expect effective mass ratio.

It can be concluded that the methods of Sader {\it et al.\ }and Cleveland {\it et al.\ }provide an effective mass ratio for a string on the same order of that calculated using Eq.~\eqref{effmass1D}, however the approximate values deviate significantly from the exact ones. For this reason, the EMI should be used to calculate the effective mass of a string whenever possible, especially when precise results are important.

\begin{table}
\centering
\begin{tabular}{ lc }
\hline \hline
Method & $m_{{\rm eff},1}/m$ \\ \hline
EMI -- Analytical & 0.5000 \\
EMI -- FEM & 0.4969 \\
Sader {\it et al.\ }-- FEM & 0.4334 \\
Cleveland {\it et al.\ }-- FEM & 0.4265 \\
\hline \hline
\end{tabular}
\caption{Values of $m_{\text{eff},n}/m$ for the fundamental mode of a simple string calculated using the different methods detailed in Sec.~\ref{effmasssec}.}
\label{stringfundtab}
\end{table}

\subsection{Effective masses of membranes}

\subsubsection{Rectangular membranes}

We begin by calculating the effective mass ratio for a rectangular membrane analytically by inputting Eq.~\eqref{sqmembmodesh} into the EMI giving

\begin{equation}
\frac{m_{{\rm eff},{mn}}}{m} = \frac{1}{L_x L_y} \int_0^{L_y} \! \int_0^{L_x} \! dx \, dy \, \left[ \psi_{mn} (x,y) \right]^2 = \frac{1}{4}.
\label{anmassrecmem}
\end{equation}

\noindent This result is very simple and mode independent, as with the strings. In fact, we get the value that one would expect from multiplying the effective mass ratio for two strings, which is unsurprising as the mode shape found in Sec.~\ref{antheories} was equivalent to multiplying the mode shapes of two strings. We also evaluate the EMI for a rectangular membrane using FEM simulation, the results of which are summarized in Table \ref{recmembmulttab}. It can be seen that these values agree very well with the analytically calculated value of 1/4.

\begin{table}
\centering
\begin{tabular}{ cc }
\hline \hline
Mode Label &  $m_{{\rm eff},mn}/m$ \\
$(m,n)$ & EMI -- FEM\\ \hline
(1,1) & 0.2498 \\
(2,1) & 0.2498 \\
(3,1) & 0.2498 \\
(1,2) & 0.2498 \\
(2,2) & 0.2498 \\
(4,1) & 0.2498 \\
(3,2) & 0.2498 \\
(1,3) & 0.2498 \\
(4,2) & 0.2498 \\
(5,1) & 0.2498 \\
\hline \hline
\end{tabular}
\caption{Values of $m_{{\rm eff},mn}/m$ calculated by solving the EMI using FEM for the first ten modes of a rectangular membrane. The mode indices $(m,n)$ match that of Fig.~\ref{rectmembmodesh}.}
\label{recmembmulttab}
\end{table}

\subsubsection{Circular membranes}

As with the rectangular membranes, we begin by calculating the effective mass ratio of a circular membrane analytically, inputting Eq.~\eqref{circmembmode} into the EMI, which gives

\begin{equation}
\frac{m_{{\rm eff},mn}}{m} = \frac{1}{\pi a^2} \int_0^{2 \pi} d \phi \int_0^{a} \! s \, ds \, \left[ \psi_{mn} (s,\phi) \right]^2.
\label{anmasscircmem}
\end{equation}

\noindent This integral can be performed for both the cases $m=0$ and $m>0$ \cite{gradshteyn} and, using the boundary condition that $J_m(\alpha_{mn}) = 0$ along with the Bessel function relation $J_{m-1}(s) + J_{m+1}(s) = 2mJ_m(z) / z$ \cite{morse}, we get the following effective mass ratios for circular membranes:

\begin{equation}
\frac{m_{{\rm eff},mn}}{m} = 
\begin{dcases}
\left[{J_1}(\alpha_{0n})\right]^2 & \text{if $m = 0$}, \\
\frac{K_m}{2} \left[J_{m+1}(\alpha_{mn})\right]^2 & \text{if}~m > 0.
\end{dcases}
\label{anmasscircmemfin}
\end{equation}

Again, the EMI can also be solved using FEM. The effective mass ratio for a circular membrane calculated both analytically and using FEM simulation are presented in Table \ref{circmembmulttab}. Upon comparing these values, it is evident that the analytical and FEM calculation match very closely.

\begin{table}
\centering
\begin{tabular}{ ccc }
\hline \hline
Mode Label & \multicolumn{2}{c}{$m_{{\rm eff},mn}/m$} \\ \cline{2-3}
$(m,n)$ & EMI -- Analytical & EMI -- FEM \\ \hline
(0,1) & 0.2695 & 0.2693\\
(1,1) & 0.2396 & 0.2394 \\
(2,1) & 0.2437 & 0.2435\\
(0,2) & 0.1158 & 0.1188\\
(3,1) & 0.2357 & 0.2356\\
(1,2) & 0.1330 & 0.1329\\
(4,1) & 0.2254 & 0.2253\\
(2,2) & 0.1556 & 0.1555\\
(0,3) & 0.0737 & 0.0753\\
(5,1) & 0.2152 & 0.2151\\
\hline \hline
\end{tabular}
\caption{Values of $m_{{\rm eff},mn}/m$ calculated by solving the EMI both analytically and using FEM for the first ten modes of a circular membrane. The mode indices $(m,n)$ match that of Fig.~\ref{circmembmodesh}.}
\label{circmembmulttab}
\end{table}

We note that, due to their poor results in calculating the effective mass ratio for a nanostring, the methods of Sader {\it et al.} and Cleveland {\it et al.} were not used to calculate the effective mass ratios for membranes, as the systems are analogous.

\section{Torsional Resonators}

Torsional nanomechanical resonators are very useful devices, as they are able to measure applied torques with excellent precision \cite{kim,losby,zhang,hchan,chabot,kleiman}. By accurately calibrating these devices, studies of many physical phenomena, such as electric charge \cite{clelandpaper}, magnetism \cite{davis,burgess,burgess2}, and superfluidity \cite{bishop}, can be explored. In this section, we will investigate how to properly calibrate such devices by using the procedure described above.

Difficulties often arise when attempting to perform thermomechanical calibration of a torsional resonator, as in most cases its motion cannot be described by a one-dimensional model. Asymmetries of the device \cite{davis} may call for a model which extends in a third dimension to describe the device's geometry or may cause hybridization of torsional modes with other degrees of freedom, such as flexural modes \cite{kim}. However, using the formalism described above, we can easily deal with such cases by the using the EMI, as it can be used to calculate the effective mass corresponding to any motion. This effective mass can then be used to properly calibrate the device. 

It may seem awkward to attribute an effective mass (linear spring constant) to a torsional resonator, as we would expect that it would be described by an effective moment of inertia (torsional spring constant). However, what needs to be remembered is that the definition of effective mass is arbitrary, and we have chosen in this paper to define it in such a way that it can be applied to any resonator. Therefore, it is possible to thermally calibrate the motion of a torsional resonator using an effective mass. We must be careful here though, as the effective mass defined by the EMI corresponds to the displacement of the resonator and not its angular motion. Only once we have properly calibrated the device's displacement will we be able to deduce its angular motion. It is important to note that we could have alternatively chosen to calibrate the angular motion first and then use it to determine the device's displacement, but such a method differs from the calibration scheme described in this document.

\begin{figure}
\includegraphics[width=\columnwidth]{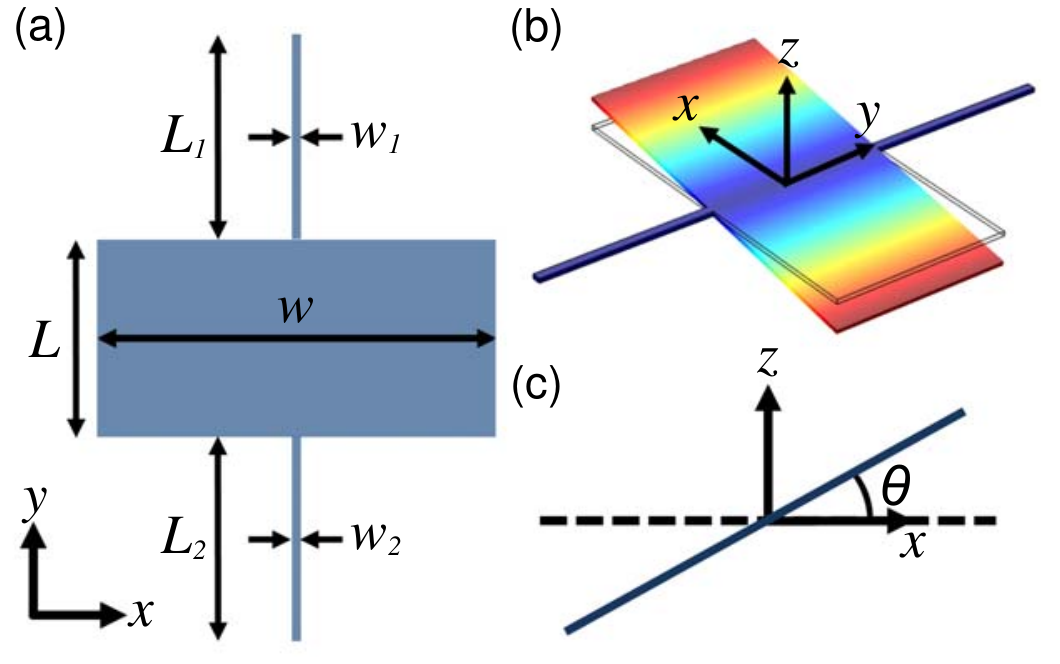}
\caption{(a) Torsional device viewed from the negative z-axis. (b) Deflected mode shape for the fundamental torsional mode. Colorscale indicates displacement with blue as zero displacement and red as maximum displacement. (c) Fundamental mode of the torsional device as viewed from the postive y-axis. The dashed line respresents the device's equilibrium position.}
\label{torsfigdim}
\end{figure}

To demonstrate how the calibration method described in this paper can be used to calibrate the angular motion of a torsional device, we present the thermal calibration for the fundamental mode of the simple torsional resonator depicted in Fig.~\ref{torsfigdim} as a case study. We will first calibrate its angular motion directly, then show how the exact same results can be achieved using the calibration method detailed above. 

Due to the symmetry of this device, only the angle of deflection $\theta(t)$ must be known in order to fully describe the motion of its fundamental mode, which is given by

\begin{equation}
z_1(x,t) = x \tan\theta(t) \approx x \theta(t),
\label{torsmot}
\end{equation}

\noindent where $x$ and $\theta(t)$ are those defined in Fig.~\ref{torsfigdim}. Our utilization of the small angle approximation in Eq.~\eqref{torsmot} is justified by the fact that for thermal motion, the deflection angle is extremely small \cite{kim}. As our resonator exhibits oscillatory motion, we model $\theta(t)$ as a damped harmonic oscillator given by \cite{wang}

\begin{equation}
\ddot{\theta} + \frac{\omega_1}{Q} \dot{\theta} + {\omega_1}^2 \theta = \frac{\tau(t)}{I_{\rm eff}},
\label{torsharm}
\end{equation}

\noindent where $\omega_1$, $Q$, and $I_{\rm eff}$ are the angular frequency, quality factor, and effective moment of inertia for the fundamental mode of this device and $\tau(t)$ is a time dependent torque applied to the system due to thermal fluctuations. We can determine the potential energy for the fundamental mode of the resonator:

\begin{eqnarray}
\label{torspot}
U &=& \frac{1}{2} \omega_1^2 \rho A \int \! dx \, \left| z_1(x,t) \right|^2 \\ 
&=& \frac{1}{2} \omega_1^2 \left|\theta(t) \right|^2  \rho A \int \! dx \, x^2 \\
&=& \frac{1}{2} \omega_1^2 \left| \theta(t) \right|^2 I_{\rm eff},
\end{eqnarray}

\noindent where we have defined the effective moment of inerta to be $I_{\rm eff} = \rho A \int \! dx \, x^2$, which turns out to be normal definition of moment of inertia in this case. Note that in the above equation we have made the approximation that the vast majority of the potential energy of the resonator will be due to the paddles of the torsion device and not the torsion rod itself. This approximation is valid for this device, as the moment of inertia of the paddles dominate that of the torsion rod. Applying the equipartition theorem to the above equation gives

\begin{equation}
\langle U \rangle =  \frac{1}{2} I_{\rm eff} {\omega_1}^2 \langle \theta^2(t) \rangle = \frac{1}{2} k_B T.
\label{equiparttors}
\end{equation}

Upon inspection of Eqs.~\eqref{torsharm}, \eqref{torspot}, and \eqref{equiparttors}, we see that is it identical to the corresponding equations in Sec.~\ref{psdsec} if we take $m_{{\rm eff},n} \rightarrow I_{\rm eff}$, $f(t) \rightarrow \tau(t)$, $a_n(t) \rightarrow \theta(t)$, and $\omega_n \rightarrow \omega_1$. Therefore, the thermal PSD for its angular motion is
\begin{equation}
S_{\theta \theta}(\nu) = \frac{k_B T \nu_1}{2 \pi^3 I_{\rm eff} Q \left[ \left( \nu^2 - {\nu_1}^2 \right)^2 + \left( \nu \nu_1 / Q \right)^2 \right]},
\label{PSDTfI}
\end{equation}

\noindent or alternatively by defining an effective torsional spring constant as $\kappa_{\rm eff} = {\omega_1}^2 I_{\rm eff}$, we have \cite{losby}

\begin{equation}
S_{\theta \theta}(\nu) =  \frac{2 k_B T \nu_1^3}{\pi \kappa_{\rm eff} Q \left[ \left( \nu^2 - \nu_1^2 \right)^2 + \left( \nu \nu_1 / Q \right)^2 \right]}.
\label{PSDTfkap}
\end{equation}

\noindent We can then fit the experimentally obtained PSD data with a function of the form

\begin{equation}
S_{VV}(\nu) = S_{VV}^\text{w} + \beta S_{\theta \theta}(\nu).
\label{PSDTfitang}
\end{equation}

\noindent This equation is identical to Eq.~\eqref{PSDTfit}, except with a conversion factor $\beta$, which allows for conversion of the theoretical angular PSD (in units of ${\rm rad}^2 / \text{Hz}$) to the experimental voltage PSD (units of $\text{V}^2 / \text{Hz}$). Once $\beta$ is determined, we have all of the information we need to fully calibrate the angular motion of our data.

It is easy to show that the calibration we just performed can be obtained by using the general method that was developed in the previous sections. We begin by determining $z_1(x,t)$ for this resonator according to Eq.~\eqref{oned}, which will be given by

\begin{equation}
z_1(x,t) =  a_1(t) u_1(x) = a_1(t) \frac{2 x}{w}.
\label{onedtors}
\end{equation}

\noindent Here $w$ is the width of the torsional resonator as given in Fig.~\ref{torsfigdim}. Note that we have carefully chosen $u_1(x)$ such that $\left| u_1(x) \right| = 1$ at its location of largest displacement located at $x = \pm w/2$. The effective mass of this mode is then determined to be

\begin{equation}
m_{{\rm eff},1} = \rho A  \int \! dx \, \frac{4x^2}{w^2}= \frac{4}{w^2} I_{\rm eff}.
\label{mefftors}
\end{equation}

We can equate Eqs.~\eqref{onedtors} and \eqref{torsmot}, to obtain a relation between $\theta(t)$ and $a_1(t)$ given by $\theta(t) = 2  a_1(t)/ w $. Using the relation between a device's motion and its PSD given by Eq.~\eqref{meanPSD}, we find that $S_{\theta \theta}(\nu)$ and $S_{zz}(\nu)$, which are the PSDs corresponding to $\theta(t)$ and $a_n(t)$, respectively, are related by

\begin{equation}
\begin{split}
S_{\theta \theta}(\nu) &= \frac{4}{w^2} S_{zz}(\nu) \\
&= \frac{4}{w^2} \frac{k_B T \nu_1}{2 \pi^3 m_{{\rm eff},1} Q \left[ \left( \nu^2 - {\nu_1}^2 \right)^2 + \left( \nu \nu_1 / Q \right)^2 \right]} \\
&= \frac{k_B T \nu_1}{2 \pi^3 I_{\rm eff} Q \left[ \left( \nu^2 - {\nu_1}^2 \right)^2 + \left( \nu \nu_1 / Q \right)^2 \right]}
\end{split}
\label{PSDrelate}
\end{equation}

\noindent This gives us the exact same $S_{\theta \theta}(\nu)$ as we obtained calibrating the angular motion directly. This tells us that we can equivalently describe the motion of this torsional resonator using an effective mass or an effective moment of inertia. The only difference between the two is that we are measuring a different degree of freedom in each case, angular for the effective moment of inertia and translational for the effective mass.

We can also analyze our simple torsional resonator by examining its effective mass ratio, as well as its effective moment of inertia ratio $I_{\rm eff} / I$ where $I = \int \! dV \, \rho ({\bf r}) {\bf r}^2 = \rho A \int \! dx \, x^2$ is the conventional moment of inertia, by using methods similar to those used in Sec.~\ref{effmasssec}. Analytically calculating the effective mass ratio for this torsional resonator by inputting Eq.~\eqref{onedtors} into Eq.~\eqref{effmass1D} produces

\begin{equation}
\frac{m_{\text{eff},n}}{m} = \frac{1}{w} \int_{-w/2}^{w/2} \! dx \, \left| u_1(x) \right|^2 = \frac{1}{3}.
\label{effmassantors}
\end{equation}

\noindent From this result, we can determine the effective moment of inertia for this system by using Eq.~\eqref{mefftors}. Dividing this result by $I=mw^2/12$, which is the conventional moment of inertia for the rectangular paddle of the torsional resonator, gives $I_{\rm eff} / I = 1$ for this device.  This result is expected because, as mentioned before, in this situation the effective moment of intertia is defined to be identical to the conventional moment of inertia. Both the effective mass and moment of inertia ratios can also be calulated using FEM simulation, and are compared to the analytical values in Table \ref{torsfundtab}.

We can also approximate the effective mass and moment of inertia ratios of our resonator by adapting the methods of Sader {\it et al.\ }and Cleveland {\it et al.\ }to be used on a torsional device. For the method of Sader {\it et al.}, we apply a known torque $\tau_0$ to the resonator and measure its angular displacement $\theta_0$, which allows us to determine the torsional spring constant via Hooke's Law, $\tau_0 = -\kappa \theta_0$. We can then approximate the effective moment of inertia using the relation $I_{\rm eff} = \kappa / {\omega_1}^2$. To perform these calculations, a FEM static solver is used, in which a force is applied to the two edges of the torsion paddle, equal in magnitude and opposite in direction, providing a torque and an angular displacement. This allows us to determine $\kappa$, which we combine with $\omega_1$ obtained from FEM eigenfrequency analysis to calculate $I_{\rm eff}$ using this method. The effective mass corresponding to this method can also be determined using Eq.~\eqref{mefftors}. The results for both the effective mass and moment of inertia ratios using the method of Sader {\it et al.\ }are summarized in Table \ref{torsfundtab}.

The method of Cleveland {\it et al.\ }can also be adapted for use on torsional resonators. In the analogous process, we observe the device's shift in resonance frequency due to an added moment of inertia $I_{\rm add}$, which is described by

\begin{equation}
\nu = \frac{1}{2 \pi} \sqrt{\frac{\kappa_{\rm eff}}{I_{\rm eff} + I_{\rm add}}}.
\label{addmom}
\end{equation}

\begin{figure}
\includegraphics[width=\columnwidth]{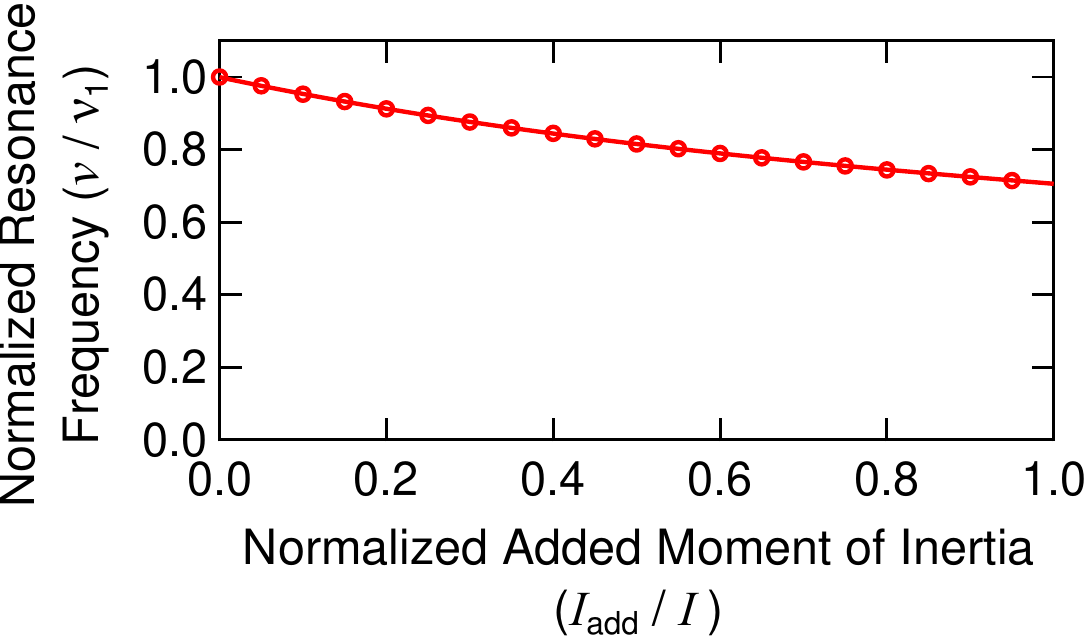}
\caption{Plot of normalized frequency versus normalized added moment of inertia for the fundamental mode of the torsional resonator depicted in Fig.~\ref{torsfigdim}. The frequency values are normalized by dividing by the fundamental frequency of each device with $I_{\rm add}=0$. The added moment of inertia is normalized by dividing by the conventional moment of inertia $I$ of the device. Solid lines are fit to Eq.~\eqref{addmom}. The effective moment of inertia ratio extracted from the fit is $I_{\rm eff}/I=0.9948$, which corresponds to an effective mass ratio of $m_{{\rm eff},1}/m=0.3316$. Both of these results can be found in Table \ref{torsfundtab}.}
\label{addmasstorsfig}
\end{figure}

\noindent To determine this frequency shift, we perform FEM simulation on the device, in which mass is added to the resonators paddles, providing an increased moment of inertia. Simulation data for such a procedure is presented in Fig.~\ref{addmasstorsfig}, along with a fit to Eq.~\eqref{addmom}, which allows us to extract an effective moment of inertia ratio and, via Eq.~\eqref{mefftors}, an effective mass ratio, which are given in Table \ref{torsfundtab}.

By investigation of Table \ref{torsfundtab}, we can see that the methods of Sader {\it et al.\ }and Cleveland {\it et al.\ }provide a very good approximation of the effective moment of inertia and mass. However, it should be mentioned that while these methods work well for simple torsional devices, such as the one presented here, they quickly break down as we move to hybridized modes, which have both torsional and flexural oscillations. For this reason, it is best to use the EMI to calculate an effective mass associated with a torsional device, and use it to calibrate the motion of the system.

\begin{table}
\centering
\begin{tabular}{ lcc }
\hline \hline
Method & $m_{{\rm eff},1}/m$ & $I_{\rm eff}/I$\\ \hline
EMI -- Analytical & 0.3333  & 1.0000\\
EMI -- FEM & 0.3314 & 0.9942 \\
Sader {\it et al.\ }-- FEM & 0.3310 & 0.9929 \\
Cleveland {\it et al.\ }-- FEM & 0.3316  & 0.9948 \\
\hline \hline
\end{tabular}
\caption{Values of $m_{{\rm eff},1}/m$ and $I_{\rm eff}/I$ for the fundamental mode of the torsional resonator depicted in Fig.~\ref{torsfigdim} calculated using the different methods detailed in Sec.~\ref{effmasssec}.}
\label{torsfundtab}
\end{table}

This simple example demonstrates that it is possible to thermally calibrate the motion of a torsional resonator using an effective mass, provided that we keep track of which degree of freedom we are measuring. This is an important result as it reinforces the fact that the method described in the previous sections can be used for any arbitrary oscillatory motion, be it flexural, torsional, or a complex combination of both.

\section{Conclusion}

This document provides a general procedure by which the thermal motion of any nano/micro-mechanical resonator can be properly calibrated via the equipartition theorem. It was shown that for any given resonator, a generic spectral density function can be used to describe its motion in the frequency domain. The only modification existing between different structural geometries was manifested by a single parameter, effective mass, which is a mode shape dependent quantity. Therefore, the effective mass was investigated in great detail for a number of nano/micro-resonator geometries, including cantilevers, doubly clamped beams, strings, membranes and torsional devices. The end result was a universal method by which any nano/micro-mechanical resonator can be thermomechanically calibrated, regardless of the complexity of its motion or number of degrees of freedom. By applying this method, the motion of such devices can be very precisely determined, allowing for use of nano/micro-mechanical resonators for a number of different measurement applications, as well as in the investigation of fundamental scientific phenomena, with confidence and accuracy.

\vspace{0.5cm}
\acknowledgments

The authors wish to thank University of Alberta, Faculty
of Science; the Canada Foundation for Innovation; the Natural Sciences and Engineering Research Council, Canada;
and Alberta Innovates Technology Futures for their support of
this research. We would also like to thank Wayne Hiebert for
helpful discussions.

\end{document}